\documentclass[11pt]{article}
\usepackage{amsfonts}
\usepackage{graphicx}
\usepackage{amsmath}
\usepackage{amssymb}
\usepackage{geometry}

\input{tcilatex}

\begin{document}

\ \ \ \ \ \ \ \ \ \ \ \ \ \ \ \ {\LARGE \ }

{\LARGE Black magic session of concordance: }

{\LARGE Regge mass spectrum from Casson's invariant}

\bigskip

\ \ \ \ \ \ \ \ \ \ \ \ \ \ \ \ \ \ \ \ \ \ \ \ \ \ \ Arkady Kholodenko

\bigskip\ \ \ \ \ \ \ \ \ \ \ \ \ \ \ \ \ \ \ \ 375 H.L. Hunter Laboratories,

\ \ \ Clemson University, Clemson, SC 29634-0973, USA{\LARGE \bigskip
\bigskip\ }

\ \ \ \ \ \ \ \ \ \ \ \ \ \ \ \ \ \ \ \ \ \ \ \ \ \
string@clemson.edu\medskip

Recently, there had been a great deal of interest in obtaining and
describing of all kinds

of knots in links in hydrodynamics, electrodynamics, non\ Abelian gauge
field \ theories

and gravity. Although knots and links are observables of the Chern-Simons
(C-S)

functional, the dynamical conditions for their generation lie outside of the
scope of the

C-S theory. The nontriviality of dynamical generation of knotted structures
is caused

by \ the fact that the complements of all knots/links, say, in $S^{3}$ are
3-manifolds which

have positive, negative or zero curvature. The ability to curve the ambient
space thus

far is attributed to masses. The mass theorem of general relativity requires
the ambient

3-manifolds to be of non negative curvature. Recently,we established that,
in the

absence of boundaries, complements of dynamically generated knots/links are

represented by 3-manifolds of non negative curvature. This fact opens the
possibility

to discuss masses in terms of\ dynamically generated knotted/ linked
structures. \ The

key tool is the notion of knot/link concordance. The concept \ of
concordance is a

specialization of the concept of cobordism to knots and links. \ The logic of

implementation of the concordance concept to physical masses results in new

interpretation of Casson'surgery formula in terms of\ the Regge
trajectories. The

latest thoroughly examined Chew-Frautschi (C-F) plots associated with these

trajectories demonstrate that the hadron mass spectrum for both mesons and

baryons is nicely described by the data on the corresponding C-F plots. \
The physics

behind \ Casson's surgery formula is similar but not identical to that
described

purely phenomenologically by Keith Moffatt in 1990. The developed topological

treatment is fully consistent with available rigorous mathematical and
experimentally

observed results related to\ physics of\ hadrons.\medskip


theory of knot and link concordance; theory of 3 and 4-manifolds;
Chew-Frautschi plots;

Regge trajectories.

\bigskip


\bigskip \bigskip

{\Large 1. Introduction}

\bigskip

Although the discovered in 2012 Higgs boson is believed \ to be \ supplying
masses to all known particles, already in 2011 there appeared a remarkable
paper \ by Atiyah et al $^{1}$ in which \ the topological origin of masses
of stable particles was discussed. In addition, in 2014, in the paper by
Buniy et al$^{2}$ the alternative model topologically generating the
glueball mass spectrum was discussed. \ Much earlier, the Abelian reduction
of QCD developed by Faddeev and Niemi$^{3}$ and subsequently by many other
authors resulted in the Faddeev-Skyrme (F-S) model whose stable
configurations are Hopfions$^{4}$. \ Hopfions are knotted/linked stable
configurations obtained by minimization of the F-S model functional.
Hopfions are believed to be \ responsible for the glueball mass spectrum of
QCD$^{5}$. According to Baal and Wipf $^{6}$, and also to Langmann and Niemi$%
^{7}$, and Cho$^{8}$, Hopfions are believed to be \ stable \textit{vacuum}
configurations while the "instantons are viewed as configurations that
interpolate between different knotted vacuum configurations"$^{7}$. Such a
vision is not shared by the authors of$^{2}$. Instead, they claim that
glueball spectrum should be associated with various types\footnote{%
That is non vacuum (or ground state).} of excitations of the Yang-Mills \
(Y-M) fields. In both cases the glueball spectrum was obtained by completely
ignoring the quark masses. Thus in all three cases, just described, the
masses were generated topologically but mechanisms of mass generation in all
these cases are profoundly different. This is unsatisfactory. Furthermore,\
while the mass-generating mechanism of $^{1}$ requires fundamental
reconsideration of the whole existing formalism of quantum fields, the
proposal based on the F-S model, in principle, allows to develop known QCD
formalism rather substantially$^{9}$. This fact \ leaves us no hope for
finding a bridge between the F-S model and that suggested in $^{2}$.
Naturally, the ultimate judge is the experiment. The authors of both$^{2}$
and$^{5}$ were comparing their theoretical results against
experimental/theoretical results of$^{10}$. Unfortunately, the results of $%
^{10}$ were subsequently criticized in$^{11}$. Not surprisingly, the
existing experimental review papers, e.g. read$^{12}$, do not mention at all
theoretical results of either$^{2}$ or$^{5}$ \footnote{%
The results of $^{2}$ are the latest in the series of papers by the same
authors written before$^{12}$ was published.}. Thus the bulk of experimental
work on the glueball mass spectrum was not guided thus far by the
predictions of \ the existing theoretical models just described. \ On the
theoretical side, there are also very serious difficulties with both models.
Indeed, the existing rigorous mathematical treatments of the Y-M fields are
based on works by Andreas Floer. His contributions are discussed in detail
in monographs by Donaldson$^{13}$ and by Kronheimer and Mrowka$^{14}$. A
brief summary of Floer's ideas is given in our work$^{15}$. \ Incidentally,
Floer's work was inspired by the work of Clifford Taubes on the Casson
invariant and gauge theory$^{16}$ $\footnote{%
The latest simplification of Taubes work is developed by Masataka$^{17}$ .}$.

In this work we develop a connection between theoretical results on the
Casson invariant and experimental results interpreted in the style of
Regge-type phenomenology. \ Our treatment does not involve any
string-theoretic formalism yet. It is solely based on known rigorous
mathematics results though. Contrary to the results of Baal and Wipf$^{6}$ ,
who believe that Hopfions are stable vacuum configurations of the Y-M
fields, in Floer's theory knots/links are \textsl{not} the ground states of
the Y-M vacua. The ground states are represented by \ 3-manifolds$^{13-16}$
. This fact eliminates the F-S type models from consideration. In the works
by Buniy et all$^{2}$ there are no instantons. Therefore, they also fall out
from consideration. There is a \ substantial number of derivations of the
Casson's invariant\footnote{%
We were able to find 16 different derivations.} of various degree of
complexity. The purpose of this work is not just to reproduce known
derivations of the Casson invariant. We have no intentions to copy this or
that mathematical result word-for-word. Instead, in many instances, being
guided by physical considerations, some new elements in deriving Casson's
invariant are obtained. This task required us to use a huge amount of facts
from knot/link theory \ scattered in literature. \ Almost all of these facts
\ have not found yet their place in the knot theory textbooks. Let alone,
the literature on knot/link theory aimed at physically educated readers is
lacking altogether this type of problematics.

\bigskip

\bigskip {\Large 2. \ }{\large Dynamically generated knotted and linked
structures.}\ \ 

\ \ \ \ \ \ {\large Review of existing results}{\Large \bigskip }\ \ 

\ \ \ \ \ \ The isomorphism between the dynamics of incompressible
Euler-type fluids and Maxwellian electrodynamics is well documented$^{15,18}$%
. In 1985 Moffatt$^{19}$ conjectured that in steady incompressible
Euler-type fluid flows \ the streamlines could have knots/links of all
types. In 2000 Etnyre and Ghrist using methods of contact geometry developed
the existence-type proof of the Moffatt conjecture$^{20}$. Subsequently, in
2012, a different type of existence-type proof was published by Enciso and
Peralta-Salas$^{21}$. Using the isomorphism between the incompressible
fluids and Maxwellian electrodynamics the constructive-type proof of
Moffatt's conjecture was developed for Maxwellian electrodynamics in$^{22}$
based on methods of contact geometry and topology$^{18}$. The proof of
Moffatt conjecture for Maxwellian electrodynamics is opening Pandora's box
of all kinds of puzzles. Indeed, since publication of works by Witten$^{23}$
and Atiyah$^{24}$ it is known that the obsevables for both Abelian and non
Abelian source-free gauge fields are knotted Wilson loops/links. It was
demonstrated that only the non Abelian Chern-Simons (C-S) topological field
theory is capable of recognizing the nontrivial knots/links. By "nontrivial"
we mean knots other than unknots, Hopf links and torus-type knots/links
which require for their description only the linking numbers and writhe(s).
Being topological in nature, the C-S functional is not capable of taking
into account the boundary conditions. In the meantime the boundary
conditions do play an important role in the work by Enciso and Peralta-Salas$%
^{21}$. \ In general, the path integral methods become impractical whenever
there is a need to take into account the non trivial boundary conditions.
E.g. everybody familiar with the path integral methods knows that even such
"simple" problem as solving the path integral for a free particle confined
into \ triangle represents a substantial challenge. The Abelian reduction of
the Y-M fields, e.g. that for the F-S model, is making them to be
describable by the Abelian-type C-S field theory. This reduction was
demonstrated in$^{7,8}$ for the Y-M fields and in$^{25,26}$ for Einsteinian
gravity formulated as gauge theory for the Lorentz group. The
constructive-type proof of the Moffatt conjecture$^{22}$ underscores the
differences between the types of knots and links the C-S field theory can
produce and can detect.

Specifically, in $^{15}$ the abelianization procedure was discussed starting
with the full non-Abelian C-S gauge field functional. This procedure differs
from that, say, described \ in $^{3}$ by the fact that it uses the Arnol'd
inequality$^{15,18}.$ Its use is equivalent to the imposition of the
Beltrami condition: $\nabla \times \mathbf{v=}\kappa \mathbf{v.}$ Here $%
\mathbf{v}$ is the Abelian gauge field \ and $\kappa $ is some nonnegative
constant. \ This condition was used \ by both Etnyre and Ghrist$^{20}$ and
by Enciso and Peralta-Salas$^{21}$. The account for the boundary conditions
\ in respective papers was done differently though. In$^{21}$ the boundary
condition \ was chosen as: \textbf{v}$\mid _{\Sigma }=\mathbf{w.}$ Here $%
\Sigma $ is embedded oriented analytic surface in \textbf{R}$^{3}$ while 
\textbf{w} is the vector tangent to $\Sigma .$ In $^{20}$ the account was
made of the fact that the Beltrami condition admits interpretation in terms
of contact geometry and topology$^{18}$. While the symplectic geometry is
used for description of dynamics on even dimensional manifolds (e.g. recall
the phase space of classical mechanics), the contact geometry is operating
in spaces of odd dimensionality. Clearly, \textbf{R}$^{3}$ is such a space.
The one point compactification converts it into $S^{3}$. Known isomorphism
between the classical mechanics and the hydrodynamics of Euler-type
icompressible fluids$^{18}$ allows us to relate the question about the
existence of closed orbits on constant energy surfaces (e.g. on S$^{3})$ to
the solution of Weinstein conjecture. This conjecture was solved by Taubes,
the same person who obtained the Casson invariant via gauge theory$^{16}$.
Not surprisingly, his solution also involves use of the Seiberg-Witten and
Floer theories$^{27}$. \ The Etnyre-Ghrist solution involves uses of special
universal template instead. It will be described below. Its use is
equivalent to use of the boundary conditions in$^{21}.$ Because of this,
both solutions are of existence-type. If the boundary conditions are
disregarded, the dynamics of closed orbits on $S^{3}$ becomes strictly
Hamiltonian and all closed orbits for such a case were classified in the
paper by Fomenko and Zung$^{28}$. These results were reobtained in our work$%
^{22}$ with help of different type of methods. The main result of the \
Fomenko-Zhung (F-Z)paper can be formulated as \medskip

\textbf{Theorem} \textbf{2.1. a) }(Fomenko-Zung$^{28}$) \textit{Generalized
iterated torus knots are precisely all possible links of stable periodic
trajectories of integrable systems on} $S^{3}.$ \medskip

This theorem can be conveniently restated as follows \medskip

\textbf{Theorem 2.1.b)}\textit{\ Generalized iterated torus knots are knots
obtained from trivial knots by toral windings and connected sums operations.
These are the only knots/links of stable }

\textit{periodic trajectories of integrable systems on} $S^{3}\bigskip $

\textbf{Corollary 2.2. }The above Theorem implies that not every link of
stable periodic trajectories can be generated by some integrable dynamical
system living on $S^{3}.$ For instance,

there are no dynamically generated knots/links containing the figure eight
knot. This fact \ immediately excludes results of Etnyre and Ghrist$^{20}$
and Enciso and Peralta-Salas$^{21}.$

But these results involve account of the boundary conditions and, because of
this, they cannot be immediately compared with those by\ F-Z! Evidently, the
F-Z knots/links are exactly

those which are observables for the Abelian version of the C-S functional.
In view of the existing abelianization procedures for the Y-M and gravity
fields mentioned above,

it follows then that the totality of such abelianized fields is described by
the F-Z theorem.

\bigskip \bigskip

The physical content of this corollary will be discussed in this section
further below. In the meantime, it is of interest to relate the F-Z results
to those by Etnyre and Grist and by Enciso and Peralta-Salas. Notice, in all
three cases we are dealing with the Abelian gauge fields! \ But the presence
of boundaries \ creates all kinds of knots/links out of Abelian fields! How
this result should be understood? Notice, that solution of Moffat's
conjecture developed in$^{20,21}$ does not involve uses of \ traditional
tools of knot theory such as Alexander or Jones polynomials, etc. It does
not involve uses of knot Floer topology and so on. \ And yet, the solution
of this conjecture implies that all types of knots/links can be generated
dynamically. \ Clearly then, the question arises: How one can be sure that,
indeed, all knots and links can be generated? To answer this question we
need to discuss briefly work by Birman and Williams$^{29}$ done in 1983.
These authors posed and solved the following dynamical problem. They studied
periodic orbits in the Lorenz system. This dynamical system emerges as
finite-dimensional reduction of the Navier-Stokes equation and is made out
of three coupled ordinary differential equations. The system of equations is
not of Hamiltonian type (that is, it is dissipative) and exhibits strange
attractor, chaos, etc. The study of periodic orbits was greatly facilitated
\ by the template construction. A template $\mathcal{T}$ is a compact
branched two-manifold with boundary built from finite number of branch
charts. In short, the template is working as some kind of a switch
regulating flow. E.g. imagine some bug crawling on the figure 8. Each time
it reaches the crossing, it should decide which way to go. This is just the
simplest example of finite state automaton. In the same paper the following
conjecture was formulated\medskip

\textbf{Conjecture 2.3.} (Birman-Williams$^{29}$ ) \textit{There are no
universal templates}. \textit{That is say, there are no dynamical systems
whose closed orbits have knots and links of all types}

\bigskip

But solution of Moffatt's conjecture does present a counterexample to the
Birman-Williams conjecture! Thus, it should be somehow linked with uses of
(perhaps different) templates. Indeed, the paper by Etnyre and Ghrist$^{20}$
does involve use of different (the so called \textsl{universal) template}.
By design, such a template can support knots and links of \textsl{any} type.
Its discovery has its origin in other works though. These are having
physical significance which was left unnoticed. This deficiency is corrected
in $^{22}.$ Chronologically, the discovery of the universal template \ was
made earlier by Ghrist$^{30}.$ Subsequently, other universal templates were
constructed. \ In his paper Ghrist was guided by yet another paper by Birman
and Williams,$^{31}$ also done in 1983, whose content was linked with
results of Etnyre and Grhrist and Enciso and Peralta-Salas in $^{22}.$ In
Birman-Williams paper $^{31}$ the authors discussed knotted magnetic field
configurations surrounding a piece of wire coiled in the shape of figure 8
knot in which current flows. They demonstrated that these knotted/ linked
configurations contain knots/links of any type. Ghrist$^{30}$ streamlined
this result by designing the universal template explicitly. This result can
be connected with that of Enciso and Peralta-Salas. For this purpose one
should take into account both the Beltrami and the boundary conditions for
static magnetic field configurations. These are: \textbf{v}$\mid _{\Sigma }=%
\mathbf{w}$ and $\nabla \times \mathbf{v=}\kappa \mathbf{v.}$ Here $\Sigma $
is the surface of the wire coiled into the shape of the figure 8 knot. As it
was argued in $^{18}$ and enforced in, $^{22}$ the conditions imposed by
Enciso and Peralta-Salas are those used for superconductors. The
correspondence between the physics of incompressible Euler fluids and
physics of superconductors was discovered by Fr\"{o}lich in 1966 \ but was
left unnoticed to our knowledge. It was brought to spotlight in $^{18}.$Once
we know how to generate knots/links of all kinds, even for the Abelian gauge
fields, the following set of problems emerge.

$\mathbf{First}$, we must take into account that "knots are determined by
their complements" as demonstrated by Gordon and Luecke$^{32}$. This means
the following. Suppose we are having just one knot K , e.g. figure 8 knot,
(that is not a link!) in $S^{3}$ (we obtain $S^{3}$ by the one point
compactification of \textbf{R}$^{3}).$ The complement of K in $S^{3}$ is
3-manifold M$^{3}$with boundary. The Gordon-Luecke theorem is telling us
that for knots embedded into $S^{3}$ there is one-to-one correspondence
between knots and 3-manifolds. Notice, however, that this theorem could
become invalid as soon as we add yet another knot into $S^{3}.$ In this case
we are dealing with links (even though 2 knots are disentangled!). And such
links are called \textit{boundary} links. The \textit{concordance} is
dealing, for example, with such types of situations. But, also, with many
other situations of physical interest to be discussed below.

\textbf{Second}, the 3-manifolds (with boundaries) can be subdivided into
hyperbolic, flat and spherical types (Seifert fibered). Finer classification
of 3-manifolds was developed by Thurston$^{33}$ who conjectured that the
interior of every compact 3-manifold has a canonical decomposition into
pieces which have \textit{geometric structures}. This, the so called \textit{%
geometrization conjecture}, was subsequently proved by Perelman. A quick
physically motivated introduction to his work can be found in$^{34}$. In
short, this means that such geometric structure is described in terms of a
complete locally homogenous Riemannian metric. In terms of knots and their
complements this can be stated as follows. Altogether there are only 3 types
of knots: torus knots, satellite knots and hyperbolic knots. If K$\subset
S^{3},$ then $S^{3}\smallsetminus $K has a geometric structure if K is not a
satellite knot. It has a hyperbolic structure if, in addition, K is not a
torus-type knot. \ 

\textbf{Third}, the essence of Einstein's equations of general relativity
can be summarized as follows: 
\begin{equation}
Curvature=Matter.  \tag{2.1}
\end{equation}%
That is to say, the matter is causing the initially flat space to curve.
From arguments just provided it follows that, say, the torus-type knots
could be associated with massive particles. This means, in particular, that
(complements of) knotted structures created by, say, the abelianized Y-M
fields (e.g. read Corollary 2.2) can act as masses. This statement requires
some refinements. For one thing, unlike the electromagnetism, masses
(without charges) can only attract each other so that we have to choose once
and for all between the hyperbolic, flat and spherical 3-manifolds. In
addition, since general relativity is 3+1 dimensional theory, if we use
knots as masses, then the homotopy moves could be interpreted in terms of
time evolution. Thus, we have to deal with 4-dimensional extensions of
3-dimensional theory of knots/links$^{35,36}$. Ultimately, it is this
observation which brings us to theories of cobordism and concordance.

The choice between the hyperbolic and spherical \ manifolds can be made
based on positivity of mass theorem in general relativity. An exhaustive
treatment of this topic is given in the book by Choquet-Bruhat$^{37}$. A
quick introduction into this subject can be found in lectures by Khuri$^{38}$
and Bartnik$^{39}$ . \ Thus, the positivity of mass theorem leaves us with
the option of considering only the nonhyperbolic 3-manifolds in calculations
involving 3+1 decomposition of gravity. The nonhyperbolic 3-manifolds are
originating as complements of nonhyperbolic knots, e.g. torus or iterated
torus-type knots. But these are the only knots/links which are dynamically
generated in the absence of boundaries (Theorem 2.1)! Only such types of
knots/links can be dynamically generated in electromagnetism and
hydrodynamics as described in detail in$^{22}$. Furthermore, the Abelian
reduction of the Y-M and gravity fields, discussed in \ works by many
authors mentioned above, allows us to apply the results of $^{22}$ to these
fields practically without change.

Thus, already knotted Maxwellian fields are acting as masses even though
photons are surely massless (even though the light rays are being bent by
gravity fields). Furthermore, the same knotted fields can possess charges.
In fact, based \ on arguments \ explained in detail in$^{15,22}$ massive
charges in electomagnetism can be reinterpreted in terms of the torus-type
superconducting knots. This observation \ when superimposed with results of
the Abelian reduction of both the Y-M and gravity fields removes all
difficulties (e.g. read$^{18}$, page 97) associated with the description of
extended objects and charges in classical gravity and Y-M theories.\bigskip

{\Large 3. \ }{\large Recovering Regge mass spectrum from }

\ \ \ \ \ \ {\large knot/link concordance. Fundamentals}

\bigskip

3.1. Concordance. First glimpses

\bigskip

Since space can be curved not only by masses but by knots/links as explained
in previous section, it is of interest to investigate the extent to which it
is possible to push the correspondence between masses and knots/links. We
begin with the "thermodynamic" property of masses, i.e. of their
extensivity. That is to say, at the very basic "classical" level two masses $%
m_{1}>0$ and $m_{2}>0$ obey the law of mass conservation%
\begin{equation}
m_{1}+m_{2}=m^{T},  \tag{3.1}
\end{equation}%
where $m^{T}$ $>0$ is the total mass. In knot theory the analog of this
relation is the operation of taking the \textit{connected sum} which we
would like now briefly describe. Formally speaking, suppose we have, say,
two knots K$_{1}$ and K$_{2}$ and we would like to combine them together
somehow. The result of such an operation is conventionally denoted as K$%
_{1}\#$K$_{2}.$ The execution of this operation is not as formal as the mass
addition though. This is so, because \ with a given knot K in 3-manifold M,
K=(M, K)\ one can associate three other knots$^{40}$ : the mirror image knot 
$m$K=(-M,K), the reverse $r$K=(M,-K) and the inverse -K=$rm$K=(-M,-K) knots.
In view of $^{15,22}$, the knot K and its reverse can be associated with
particles having opposite charges (electric or magnetic in view of
electromagnetic duality) while knots living in 3-manifolds and having
opposite orientations will represent the corresponding antiparticles. The
electrically neutral species would require us to use the connected sums of
the type K$\#r$K. \ The formalism could be developed in both ways: a) that
which recognizes orientations and b), that which is blind to orientations.
This is very much the same as to use the comparison between numbers in the
ring of integers or to use the mod comparison \ between numbers. In any
case, to make things simpler, we shall only consider oriented \ knots in
oriented 3-manifolds for the time being. Then, the operation of taking the
connected sum is relatively easy to define and, to avoid redundancy, we
refer to the excellent source$^{41}$, pages 20-22, for detailed description
of this operation. Once it is defined, it can be treated algebraically and,
at this level of treatment, the intricacies associated with orientation can
be temporarily suppressed. Thus we notice that the connected sum operation
for knots/links is commutative and associative, i.e.%
\begin{equation}
\text{K}_{1}\#\text{K}_{2}=\text{K}_{2}\#\text{K}_{1};\text{K}_{1}\#(\text{K}%
_{2}\#\text{K}_{3})=\text{K}_{1}\#(\text{K}_{2}\#\text{K}_{3}).  \tag{3.2}
\end{equation}%
The commutativity and associativity of the connected sum operation is not
sufficient for making \ this operation \ as a group operation on the set of
all oriented knots $\mathcal{K}$ .\ \ Thus far $\mathcal{K}$ is only a
monoid since it contains an identity element-the unknot. To convert this
monoid into a group requires some black magic thinking. First, we need to
introduce the notion of \textit{cobordism}.

Let M$_{1}$ and M$_{2}$ be oriented closed 3-manifolds. \ By definition, M$%
_{1}$ is cobordant to M$_{2}$ if there is compact oriented smooth 4-manifold
W (called \textit{cobordism} between M$_{1}$ and M$_{2})$ such that $%
\partial W=$M$_{1}\amalg (-$M$_{2}),$where $-$M$_{2}$ denotes manifold M$%
_{2} $ with reverse orientation and $\amalg $ denotes the disjoint union.
Clearly, the cobordism is an equivalence relation. It is possible now to
apply the idea of cobordism to knots. This can be done with several levels
of sophistication. For instance, let K$_{1}$ and and K$_{2}$ be some knots.
Following $^{40}$, consider now the connected sum K$_{1}\#rm$K$_{2}$ so that
effectively $-$K$_{2}=rm$K$_{2}.$ \ In addition, following$^{42}$ we
introduce a cone \^{k} over the knot K as depicted in Fig.1.

\begin{figure}[ptb]
\begin{center}
\includegraphics[scale=1.5]{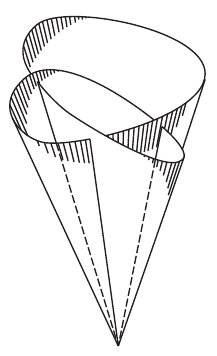}
\end{center}
\caption{ A trefoil knot bounds a non-locally flat disc. In 3 dimensions
this nonlocality is depicted as a cone with 3 boundary singularities}
\end{figure}

Topologically \^{k} is a disc D$^{2}.$ If embedded into 4-dimensional ball B$%
^{4}$, \ the intersections seen in Fig.1 will disappear. These are only
visible in $S^{3}$. The cone apex represents the only singularity of D$^{2}$
for any nontrivial K. For trivial knots this singularity disappears. The
above peculiar features of embedding of \^{k} into 4-ball can be explained
with help of the famous Whitney Theorem$^{43}$.

\textbf{Definition 3.1}. \textit{Two knots} K$_{1}$ \textit{and and} K$_{2}$ 
\textit{are smoothly concordant if there is a smooth embedding} $f$: $%
S^{1}\times \lbrack 0,1]\rightarrow S^{3}\times \lbrack 0,1]$ \textit{such
that} $f(S^{1}\times 0)=$K$_{1}$ and $f(S^{1}\times 1)=$K$_{2}.$

Concordance is the equivalence relation. It will be denoted as K$_{1}\sim $ K%
$_{2}.$

\textbf{Definition 3.2.} \ \textit{A knot} K \textit{is} \textit{slice} 
\textit{if} \ K$\sim $U, \textit{where} U \textit{is unknot}. \textit{%
Alternatively, and more formall}y, \textit{a knot being} a \textit{slice
means that it is a boundary of} D$^{2}$ \textit{smoothly} \textit{embedded in%
} B$^{4}$

It can be demonstrated, e.g. read p.88 of $^{44}$, that K$\#rm$K $\sim $U.
Use of this result allows us to define an inverse for the connected sum
operation. Indeed, the inverse for a given knot K is $rm$K. This observation
allows us to replace the monoid $\mathcal{K}$ with the \textit{concordance
group} $C_{3}.$ Physically, the combination $K\#rmK$ brings together
particle and antiparticle. They may annihilate so that the rest mass is
zero, that is U.

Now it is becoming possible to introduce some concordance invariants.
Specifically, following Murasugi $^{41}$, we notice that the knot signature $%
\sigma ($K$)$ is one of such invariants. By definition, 
\begin{equation}
\sigma (\text{K}_{1}\#\text{K}_{2})=\sigma (\text{K}_{1})+\sigma (\text{K}%
_{2})  \tag{3.3}
\end{equation}%
and $\sigma (rm$K$)=-\sigma ($K$).$Therefore \ $\sigma $(K$\#rm$K$)=\sigma ($%
K$)-\sigma ($K$)=0=\sigma ($U$).$ Thus, the signature invariant \ can detect
slice knots. Because of this property, it is also a concordance invariant.\
This means that if K is concordant to K$^{\prime }$ , then $\sigma ($K)=$%
\sigma ($K$^{\prime }).$Equation (3.3) provides a homomorphism from the
concordance group $C_{3}$ into \textbf{Z. }More accurately\textbf{, }since
the signature is always even, we have to use $\sigma ($K$)/2$ to map $C_{3}$
into\textbf{\ Z.} Are there other invariants which possess the additivity
property? \ Are they also concordance invariants? The answer to both
questions is "Yes". \ Some of them will be introduced whenever it will
become appropriate. Incidentally, by itself the Casson invariant-the main
focus of this paper- is \textsl{not} a concordance invariant. E.g. read $%
^{45}$, page XV. \ Nevertheless, as we shall demonstrate, to introduce this
invariant it is required to use many results from the concordance theory.

In the meantime, we would like to notice the following. \ It is clear that
the slices K$\#rm$K can be formally constructed for all knots. This
procedure is not without some controversy though since it ignores the
existence of the so called amphichiral knots. The figure 8 knot is the first
example of the \textit{fully amphichiral} knot. Such knots are
(ambient)isotopic with respect to both its reverse $r$ and its mirror image $%
m$. In addition, there are \textit{positive and negative amphichiral knots}
These are respectively homeomorphic to their mirror images without changing
orientation (positive) and with changing the orientation (negative). The
figure 8 knot could be identified with the Majorana fermion which is
simultaneously particle and antiparticle. Moreover the signature $\sigma ($K$%
_{8})=0$ so that figure 8 knot is formally slice. But it is not! Indeed, the
knot is slice if, in addition to its signature being zero, its Alexander
polynomial $\Delta _{\text{K}}(t)$ could be represented in the factorized
form, that is $\Delta _{\text{K}}(t)=f(t)f(t^{-1})$ with $f(t)$ being some
polynomial$^{42}$. In the case of figure 8 knot its Seifert matrix $V$ is
known to be%
\begin{equation*}
V=\left( 
\begin{array}{cc}
-1 & 1 \\ 
0 & 1%
\end{array}%
\right)
\end{equation*}%
so that its Alexander polynomial is $\Delta _{\text{K}_{8}}(t)=\det
[V-tV^{T}]=t^{3}-3t+1.$ Evidently, it is not factorizable.\ \ Furthermore,
the figure 8 knot is a hyperbolic knot. Hyperbolic knots cannot be
dynamically generated as proved in$^{22}$. This result is consistent with
the positive mass theorem in general relativity as discussed already.
Therefore, even though mathematically this type of a knot does have right to
exist, in the absence of boundaries physically it cannot be spontaneously
generated. As result, if we identify the Majorana fermion with figure 8
knot, we should admit that the Majorana fermion does not exist in Nature.
This result is valid in 3+1 dimensions. Experimentally, at the moment of
this writing, indeed , in 3+1 dimensions the Majorana fermion was not
found.\ The 2+1 dimensional version of Majorana fermion is believed to be
very valuable in theory of quantum computing $^{18}$. \ The description of
their generation in condensed matter physics lies outside of the scope of
this paper. For completeness, the current status of Majorana Fermions in
various dimensions \ is discussed in detail in $^{46}.$

To finish this section, we \ would like to quote the following\medskip

\textbf{Theorem 3.3. (}The\textbf{\ }mirror Theorem$^{44}$\textbf{, }page 173%
\textbf{). }\textit{Let}\textbf{\ }K \textit{be a knot diagram with zero
writhe (that is without curls)}. \textit{Then} K \textit{is ambient-isotopic
to} $m$K \textit{if and only if it is regularly isotopic} \textit{to} $m$%
K.\medskip

This theorem could be physically understood as follows. First, we notice
that, by definition, the regular isotopy is taking place only in 2
dimensions. Next, following Kauffman, suppose that our knot is made of
rubber tube so that twisting is causing some tension along the tube. The
twisting costs some energy associated with creation of tension. Thus to
bring knot regular-isotopically to its mirror image requires us to go over
the potential barrier associated with the creation of tension during this
transformation process. That is knot and its mirror image are sitting in two
potential wells separated by the potential barrier.

There is \ an exact analog of the situation just described in quantum
mechanics. It is described in the 3rd volume of Feynman's lectures in physics%
$^{47}$, page 8-11. Feynman is considering the ammonia molecule NH$_{3}$ as
a candidate for the working body for the ammonia maser. The ammonia molecule
\ can be visualized as a tetrahedron whose 3 vertices \ are being occupied
by H's while the 4th- by N. Consider an "up" configuration. It is a
configuration in which all 3 hydrogens are sitting on the top of the mirror
while the nitrogen is elevated above the mirror. The mirror image of such a
tetrahedron, the"down configuration", is another tetrahedron with N molecule
\ below the mirror plane. It happens, that such an ammonia molecule has an
induced dipole moment \ located at the center of symmetry of ammonia and, of
course, of its mirror image. But for the "up" configuration the dipole is
looking straight down while for the "down" configuration the dipole is
looking straight "up". Following Feynman, let $\mid 1>$ \ be the quantum
state of the "up" configuration and $\mid 2>$ \ be the quantum state for the
"down" configuration. \ Then, according to Feynman, "it is possible for the
nitrogen to push its way through three hydrogens and to flip to another
side". Such a move of N through\ H's \ is exact equivalent of regular
isotopy described by Theorem 3.3. while the tension in the present case is
associated with the energy for flipping of the dipole. As result, we just
obtained a two-level quantum mechanical system. Such a system is extensively
discussed in$^{18}$ in connection with problems of quantum computation.

There is yet another way to interpret just obtained results. It is
associated with uses of the Vassiliev invariants and virtual knots.
Following Manturov$^{48}$ , we provide the

\textbf{Definition 3.4.} A \textit{virtual link diagram is a planar
four-valent graph endowed with the following structure: each vertex \ either
has an (over) under crossing or is having crossing which is} \textit{not
resolved (a double point)}.

Virtual knots and links were discovered by Kauffman$^{49}$ and are currently
under active study.

\textbf{Definition 3.5.} ( Ref. 50, page 72]) \textit{Any knot invariant can
be extended to knots with double points by means of the} \textsl{Vassiliev} 
\textsl{skein relation. }

It is\textsl{\ } depicted in Fig.2

\begin{figure}[ptb]
\begin{center}
\includegraphics[scale=1.1]{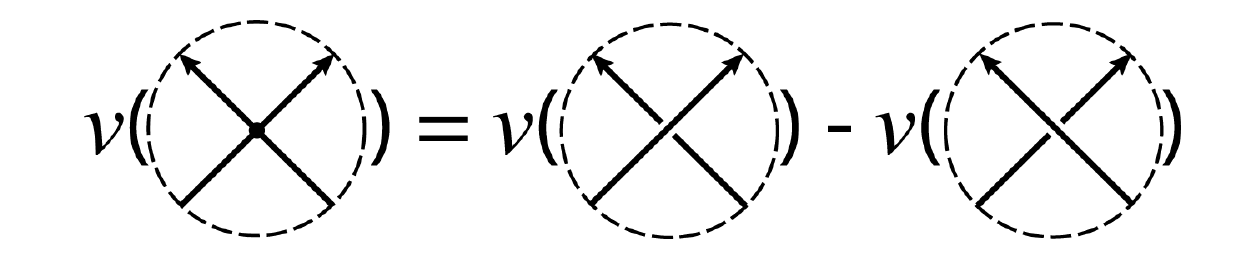}
\end{center}
\caption{ Basic skein relation for Vassilliev-type invariants}
\end{figure}

Here $\mathit{v}$ is the knot invariant with values in some Abelian group.
The left hand side is the value of $v$ on singular (that is with specific
crossing \ not resolved) knot K while the right hand side is the difference
of the values of \textit{v} on regular knots obtained from K by replacing
the double point with a positive and negative crossings respectively. \ 

The physical situation related to Theorem 3.3. is reminiscent to the theory
of spontaneous symmetry breaking during the 2nd order phase transition.
Perhaps, it is possible to develop some kind of Higgs-type mechanism of
spontaneous mass generation related to just described process. In addition,
however, there are cases when even the ambient isotopy fails to bring a
given knot K into its mirror image $m$K. In such a case we are dealing with%
\textit{\ chiral }knots whose simplest representative is trefoil. To bring
the trefoil made of rubber tube to its mirror image requires us to break the
rubber tube and to reconnect it back. This will switch an over crossing into
under crossing or vice-versa. This process surely costs energy. Physically,
it is very much the same thing as breaking a bond when a molecule wants to
escape the solid phase and go to the gas phase (and vice-versa). This is
typical case of the first order phase transition accompanied by the presence
of \ the latent heat (needed for breaking of \ chemical bonds). \ We shall
demonstrate below that in the case of \ physical interpretation of the
Casson invariant it is \ exactly these type of processes that are
responsible for generation of the Regge-type mass spectrum.

The trefoil is definitely not a slice knot but the connected sum of the
trefoil with its mirror image (known as square knot) is slice, e.g. read$%
^{51}$, page131. Although physically questionable, mathematically it is
easily possible to make the connected sum of two figure eight knots (it will
be denoted as K$_{8}^{2})$. Such a formation, when combined (via connected
sum) with yet another K$_{8}^{2}$ does yield a slice$^{51}$, page 203.
Notice as well that even though the K$_{8}$ knot is hyperbolic, K$_{8}\#$K$%
_{8}$ is not!\footnote{%
This result belongs to William Thurston. We would like to thank Morwen
Thistlethwaite, (UTK), for detailed explanation of this fact.} Questions
remain. 1. Are all fully amphichiral knots hyperbolic? \ 2. Is the same true
for positive and negative amphichiral knots? The problem of designing slice
knots/links is further complicated \ by the following observation. The
stevedore knot is known to be slice\footnote{%
In fact, it is dobly slice knot, e.g. see section \ \ 4.4 for definition and
read http://katlas.org/wiki/6\_1} but is not made as connected sum$^{44}$ K$%
\#rm$K, page 86. Furthermore, it is a hyperbolic-type knot\footnote{%
This information was supplied to us by Morwen Thistlethwaite (UTK).}. Is
there other (than stevedore) slice knots/links which are not of the form K$%
\#rm$K and \ are hyperbolic? \ Since the research on slice knots is still
ongoing, it would be too premature in this work to list all the achievements
in this domain of research. We shall restrict ourself by the most
conventional slices of the type K$\#rm$K. The \textit{ribbon} knots/links
are immediately connected with such slices and will be discussed below in
this work, e.g. see Fig.11 and read comments related to this figure.

In Section 2 we argued that the hyperbolic knots cannot be candidates for
physical masses. Fortunately, they also cannot be generated dynamically (in
the absence of boundaries). However, if the notion of dark matter does make
any sense, and it does\footnote{%
E.g. read our paper http://arxiv.org/abs/1006.4650}, it might be of some
interest to investigate what events/processes in Nature can cause hyperbolic
knots/links to come to life. Theoretically, though, this is possible only in
the presence of boundaries of special type$^{22}$ as explained in previous
section. \bigskip

3.2. From cosmological models to microcosm and back. \ 

\ \ \ \ \ \ \ How concordance with observational data brings to

\ \ \ \ \ \ \ life the Casson invariant

\bigskip

Casson's invariant was defined originally only for the homology sphere
3-manifolds. Following Rolfsen$^{42}$, let us recall the \medskip

\textbf{Definition 3.6}. \ A closed connected (but not necessarily simply
connected !) 3-manifold M is a\textit{\ homology sphere} if H$_{1}($M$)=0.$
If, in addition $\pi _{1}($M$)=1,$then the manifold is a \textit{homotopy}
sphere. \medskip

The Poincare$^{\prime }$ conjecture ( recently proven by Grisha Perelman,
e.g read Ref.$34$ for a quick introduction)\ claims \ that the 3-sphere $%
S^{3}$ is the only manifold for which both H$_{1}($M$)=0$ and $\pi _{1}($M$%
)=1$ hold. Clearly, homotopy spheres are homology spheres. Using the Poincare%
$^{\prime }$ duality it can be shown that for homology spheres all homology
groups are exactly the same as those for $S^{3}.$ Suppose now that there is
a 3-manifold such that H$_{1}($M$)=0$ but $\pi _{1}($M$)\neq 1$. If such a
manifold does exist, it represents a \textit{homology} sphere. \ Poincare$%
^{\prime }$ was the first who designed the manifold of this type known in
literature as dodecahedral space. A pedagogical account of such a space as
well as its relevance to cosmological models of Universe is discussed in \
the paper by Weeks $^{52}$. Subsequent studies put into question usefulness
of such a space as good model of Universe. \ The dodecahedral space
represents the first example of homology sphere space which is not simply
connected since its fundamental group is $\pi _{1}$ is that of binary
icosahedron.\footnote{%
E.g. read http://mathoverflow.net/questions/91760/poincare-dodecahedron-space%
} Surprisingly, This happens to be a trend. There are many homology spheres
which are not simply connected. \ Such multiply connected spaces are the
latest candidates of cosmic topological models of Universe$^{53}$. After
Poincare's description of the dodecahedral space many other designs of
homology spheres came to life. It has become possible to construct a
countable infinity of different homology spheres$^{54,55}$ some of which
will be described shortly below. Thus far the recipes for making homology
spheres \ were purely mathematical. This means that they leave \ completely
outside the question of major physical importance: \textsl{How these
homology spheres can be physically realized in nature}? Notice that the
standard knot theory \ does not contain a spatial scale. This means that
knots/links could live both in macro and microcosm and everywhere in
between. This fact brings to life many puzzles in knot theory revealing
themselves the most in the theory of 4-manifolds$^{43}$. We shall touch this
topic a bit further later in the text. In$^{15,22}$ we discussed physical
mechanisms leading to formation of knots/links. In the previous section it
was explained that in the absence of boundaries only non hyperbolic
knots/links can be formed. Complement of every knot/link, say in $S^{3},$ is
\ some 3-manifold. \ Mathematically, these 3-manifolds are obtainable, for
instance, by operation of Dehn surgery which can be reinterpreted in terms
of cobordism$^{56}$ and, hence, of concordance.\medskip

\textbf{Definition 3.7.} \textit{A knot} $K\subset S^{3}$ \textit{is said to
satisfy} \textit{property }$\ $P \textit{if there are no non-trivial
surgeries on it yielding a simply-connected manifold}.\medskip

Thus, whenever the property P holds, homology spheres are not simply
connected 3-manifolds. Incidentally, The unknot does not satisfy property P
since, for example, 1-surgery on it yields \ $S^{3},$e.g. read$^{54}$. This
result follows from the fact that $S^{3}$ can be made out of \ the union of
two solid tori (the most elementary example of Heegard splitting!). The
property P remained a conjecture till \ 2004 when it was proved by
Kronheimer \ and Mrowka$^{57}$. It is remarkable that cosmological models of
multiconnected (almost) flat Universes$^{53}$ are compatible with the
property P. Furthermore, multiconnectedness reveals itself microscopically.
We would like now to\ explain how this could happen.

We begin with the observation that the property P is satisfied \ by knots K
\ for which $\frac{d^{2}}{dt^{2}}\Delta _{\text{K}}(t)\mid _{t=1}\neq 0,$
page 662 (bottom) of Ref.58, and $^{45}$, page XV. Here, as before, $\Delta
_{\text{K}}(t)$ is the Alexander polynomial for some knot K. The combination 
$\frac{d^{2}}{dt^{2}}\Delta _{\text{K}}(t)\mid _{t=1}$enters the definition
of the Casson invariant as explained below and, since the Casson invariant
admits physical interpretation, this means that property P reveals itself
microscopically as well.

To proceed, we have to define several additional concepts e.g. that of the $%
spliced$ $sum$ \footnote{%
Although this operation is not limited to the homology spheres, in this work
we shall introduce it only in the context of homology spheres.}. Let K$_{1}$
and K$_{2}$ be oriented knots living in oriented homology spheres $\Sigma
_{1}$ and $\Sigma _{2}$ respectively. Let M$_{1}=\Sigma _{1}\setminus $K$%
_{1} $ and M$_{2}=\Sigma _{2}\setminus $K$_{2}$ be their complements. \ Both
M$_{1}$ and M$_{2}$ are manifolds with boundary: $\partial $M$_{1}$= $%
\partial $M$_{2}=T^{2},$ where $T^{2}$ is a torus. Let $(m_{1},l_{1})$ and $%
(m_{2},l_{2})$ be the canonical meridian and longitude respectively\footnote{%
In the sence of Rolfsen $^{54},$ page 136. See also Appendix.} on $\partial $%
M$_{1}$ and $\partial $M$_{2},$ then we obtain the following

\textbf{Definition 3.8.} \ \textit{The} $\mathit{spliced}$ $\mathit{sum}$ (%
\textit{along }K$_{1}$ \textit{and} K$_{2})$ \textit{of} $\Sigma _{1}$ 
\textit{and} $\Sigma _{2}$ \textit{is the operation of creating a new
homology sphere} $\Sigma =$M$_{1}$ $\cup _{T^{2}}$ M$_{2}$ \ \textit{via
gluing} M$_{1}$ \textit{and} M$_{2}$ \textit{via orientation -reversing} 
\textit{homeomorphism along their common boundary} $T^{2}$ \textit{in such a
way that} $m_{1}$ \textit{is glued to} $l_{2}$ \textit{and} $m_{2}$ \textit{%
to} $l_{1}$

The connected sum is a special case of the splicing \ sum. This will become
evident upon reading. The notion of the spliced sum leads us directly to our
first example of designing the homology spheres. \ Following Rolfsen$^{54}$,
page 251, we only should replace $\Sigma _{1}$ and $\Sigma _{2}$ by two
copies of $S^{3}$ \ while keeping both K$_{1}$ and K$_{2}$ nontrivial to
obtain the homology sphere $\Sigma =$M$_{1}$ $\cup _{T^{2}}$ M$_{2}.$ In the
case of $S^{3}$ the complements $S^{3}\setminus $K$_{i},i=1,2,$ are solid
tori$^{54,56}$. For such a case, it was rigorously proven that various knots
produce countable infinity of (non -simply connected) homology 3-spheres $%
\Sigma =$M$_{1}$ $\cup _{T^{2}}$ M$_{2}.$ Since such a design of homology
spheres is mathematically plausible but physically\ not realizable, we shall
describe another design admitting physical interpretation. Doing so gives us
the opportunity to describe the first example of magical powers of
concordance. \ 

We begin with discussion\ of the paper by Gordon$^{59}$. In his paper slight
change in the rules for construction of spliced sums are described. It is
worth discussing the alternative construction of splicing and its
implications in some details. \ 

For $i=1,2$, let K$_{i}$ be a knot in homology sphere $\Sigma _{i}$ with
exterior M$_{i}$. Furthermore, let $A=\left( 
\begin{array}{cc}
\alpha & \beta \\ 
\gamma & \delta%
\end{array}%
\right) $ be a 2$\times $2 integral matrix with det$A=-1$ defining
orientation-reversing diffeomorphism $h$: $\partial $M$_{1}\rightarrow
\partial $M$_{2}$ so that the closed 3-manifold M$_{1}\cup _{h}$M$_{2}$ will
be denoted as M(K$_{1}$,K$_{2};$A) or, better, as M(K$_{1}$,K$_{2}$;$\alpha
,\beta ,\gamma ,\delta $). Since the exterior of the unknot U is (R$%
^{2}\times S^{1})\footnote{%
Indeed, the unknot U is lying is the plane \textbf{R}$^{2}$ and the loop $%
S^{1}$ from the extra 3rd dimension is winding around the unknot.}$ the
manifolds of the type M(U,K;A) are exactly those which are obtained \ by
removing the tubular neighborhood of K and sewing it back (possibly
differently).

\textbf{Remark 3.9.} Just described operation is an example of a $surgery$ $%
along$ K (more on this is given in Appendix).

In particular, if K is knot in $S^{3},$ then the homology sphere is obtained
when M(U,K;$\alpha ,\beta ,\pm 1,\delta $). \ This was the prescription for
creating homology spheres discovered by Dehn, e.g. \ read$^{54}$ \ , page
246. If $\cong $ denotes the orientation-preserving diffeomorphism, then it
follows that M(U,K;$\alpha ,\beta ,\pm 1,\delta $)$\cong $M(U,K;$\epsilon
\alpha ,1,1,0$)$\equiv $M$($K$,\alpha ).$ Here $\epsilon =\pm 1.$
Incidentally, M(U,U$;$A) is the lens space L($\gamma ,-\alpha )\cong $L($%
\gamma ,\delta ).$ This can be seen from reading of page 99 of Ref.60.

The above results can (and should) be rewritten in terms of knot
concordance. \ To avoid any ambiguity, we shall use Gordon's notations for
this purpose. Thus, we have the following\medskip

\textbf{Definition 3.10}. \textit{Closed 3-manifolds} M$^{+}$ \textit{and} M$%
^{-\text{ }}$ \textit{are} G-\textit{homology cobordant} \textit{if there
exist \ a} 4-\textit{manifold} W \textit{such that} $\partial $W$\cong $M$%
^{+}\cup $ $-$M$^{-\text{ }},$ \textit{provided that inclusions} M$^{\pm
}\rightarrow $W \textit{induce} \textit{isomorphisms} H$_{\ast }$(M$^{\pm },$%
G)$\rightarrow $H$_{\ast }$(W,G). \textit{Under such circumstances a} G-%
\textit{homology sphere} \textit{is a closed} 3-\textit{manifold} M \textit{%
such that} H$_{\ast }$(M$,$G)$\cong $H$_{\ast }$($S^{3}$,G). \medskip

\textbf{Definition 3.11.} A \textit{proper pair of knots} $C=$(W, $%
S^{1}\times I)$ \textit{\ is such a pair that} $\partial C\cong C^{+}\cup $ $%
-C^{-\text{ }},$\textit{where} $C^{+}=$(M$^{+},S^{1}\times 0)$ \textit{and} $%
C^{-}=$(M$^{-},S^{1}\times 1)$ \textit{is a} \textit{homology cobordism} (%
\textit{that is} \textit{concordance) between knots} $C^{+}$ \textit{and} $%
C^{-\text{ }}$ \textit{if} W \textit{is homology cobordism between} M$^{+}$ 
\textit{and} M$^{-}.$

\textbf{Definition 3.12}. $\mathit{A}$ \textit{knot cobordism} (that is
concordance) \textit{between knots} $C^{+}$ \textit{and} $C^{-}$ \textit{in} 
$S^{3}$ \textit{is a proper pair} $C=$($S^{3}\times I,S^{1}\times I)$ 
\textit{such that }$C^{+}\cong $($S^{3}\times 0,S^{1}\times 0)$ \textit{and} 
$C^{-}\cong $($S^{3}\times 1,S^{1}\times 1).$ Gordon proves the following$%
^{59}$ \medskip (his Theorem 3), page 163,

\textbf{Theorem 3.13.(a) }\textit{Let} $C_{1}$ \textit{and} $C_{2}$ \textit{%
be two knot concordances} (cobordisms) \textit{such that} $C_{1}^{+}=0.$%
\textit{\ Suppose that in addition} (in matrix $A)$ $\gamma =\pm 1$ \textit{%
and that either} a) $C_{2}^{+}=0$ or b) $\alpha =0.$ \textit{Then} \textit{%
the 4 manifold} $W(C_{1},C_{2};A)$ \textit{is homology cobordism between} $%
S^{3}$ \textit{and} M($C_{1}^{-}$,$C_{2}^{-};$A) \textit{with} $\pi
_{1}(W)=1.$ \textit{By attaching a copy of} \textbf{R}$^{4}$ \textit{to} W 
\textit{along} $S^{3}$ \textit{the contractible manifold} $N$ \textit{is \
obtained}. \bigskip

This allows us, following Gordon, to restate Theorem 3.13 (a)
differently\bigskip \bigskip

\textbf{Theorem 3.13.(b) }\textit{Suppose that} (in matrix $A)$ $\gamma =\pm
1.$ \textit{If} K$_{1}$\textit{\ is a slice knot, and either} a) K$_{2}$ 
\textit{is also a slice knot}, \textit{or} b) $\alpha =0$ (in matrix $A$). 
\textit{Then the 3-manifold} M(K$_{1}$,K$_{2};$A) \textit{bounds a} \textit{%
contractible manifold} $N.\bigskip $

This theorem was \ proved subsequently by many authors in various settings.
For the purposes of this work it is of interest to restate the same theorem,
still differently, in the context of \textit{Freedman's theorem on Fake Balls%
}$^{43}$, page 83.\medskip

\textbf{Theorem 3.13 (c) (}Freedman). \textit{Every homology sphere} 3-%
\textit{manifold} $\Sigma $ \textit{bounds a contractible} 4-\textit{manifold%
} W- \textit{a fake 4-ball}.\medskip

\textbf{Corollary 3.14.} \ \ Having some knot (representing a particle) in
physical vacuum (created by the slice knot(s) in $S^{3})$ changes the
Euclidean space $S^{3}=$ \textbf{R}$^{3}+$ point at infinity into homology
sphere $\Sigma $. The properties of vacuum are determined by the presence of
fake contractible 4-manifold. Since all \textit{smooth} 4-manifolds are
symplectic, they are automatically K\"{a}hler. Therefore such manifolds, at
least locally, can be looked upon as the Calabi-Yau manifolds
(alternatively, as the K3 surfaces). This fact was noticed by Fintushel and
Stern$^{61}$ (e.g. read their Corollary 1.4.) whose work is summarized in$%
^{43,62}$. Alternatively, the observed existence of homology spheres
modelling our Universe \ is reflection of the fact that the physical vacuum
state is filled with particles and antiparticles in equal amounts. This is
in contrast with results of knot theory where, in addition to "physical"
slices K$\#rm$K, other slices had been discovered (e.g. read previous
subsection).

Based on just described information, we shall proceed in a way alternative
(but equivalent) to\ that developed by Fintushel and Stern. By studying
implications of Freedman's Fake Balls theorem we would like to reveal yet
another magical properties of concordance while trying to make our
presentation as physical as possible.

Following$^{56}$, page 73, \ we notice that any knot K $\subset S^{3}$ can
be analytically described as an intersection of the sphere $S^{3}$ =$%
\{(z,w)\in \left\vert z\right\vert ^{2}+\left\vert w\right\vert ^{2}=1\}$
with 4-ball B$^{4}=\{$ $(z,w)\in \left\vert z\right\vert ^{2}+\left\vert
w\right\vert ^{2}\leq 1\}.$ Since every knot/link bounds a Seifert surface F$%
_{K}$ \footnote{%
E.g.see Fig.10.}so that K=F$_{K}\cap S^{3}$, by looking at Fig.1 we can
associate the apex of the cone \^{k} with the center of B$^{4}$ while the
points x$\in $K and x$\in $ F$_{K}$ are to be treated as points inside the
ball B$^{4}.$ Next, we assume that there exist some function $f(x)$ \ which
we consider as the Morse function for which $f$: F$_{K}\rightarrow \mathbf{R,%
}\ $as usual. The function $f(x)$ can be selected based on the requirement $%
f(x)\neq 0$ $\forall x\in $ F$_{K}.$ To take advantage of such defined
function it is convenient to change the rules a little bit. Specifically,
the extra (time) dimension can be introduced now via $S_{t}^{3}$=$\{(z,w)\in
\left\vert z\right\vert ^{2}+\left\vert w\right\vert ^{2}=t^{2}\}$ so that
the intersection F$_{K}\cap S_{t}^{3}$ at a point $p\in $ F$_{K}\cap
S_{t}^{3}$ is transverse. This means that T$_{p}$F$_{K}\oplus $T$%
_{p}S_{t}^{3}$ spans T$_{p}$B$^{4}.$ By definition, critical points $p^{i}$
\ of the function $f(x)$ are points at which $\partial f/\partial x(p^{i})=0$%
. \ If there are such points, then the standard protocol of Morse theory can
be applied. Since F$_{K}$ is compact such designed Morse function will
contain only finite number of critical points which will be assumed to be
non degenerate (that is these are well separated points). In the degenerate
case, one should use the Morse-Bott version of Morse theory. \ For a fixed $%
t $ each intersection F$_{K}\cap S_{t}^{3}$ is a curve in 3-space which is
becoming a 2 dimensional surface $S_{K}$ in B$^{4}.$ If there is a link,
say, made of two knots K$_{1}$ and K$_{2}$, then it is appropriate to talk
about the intersection of the surface $S_{K_{1}}$ with $S_{K_{2}},i.e.$ $S_{%
\text{K}_{1}}\bullet $ $S_{\text{K}_{2}}.$ The details of this construction
are beautifully explained in$^{43}$, chapter 4. The result $S_{K_{1}}\bullet 
$ $S_{K_{2}}$ can be understood on example of a Hopf link as depicted in
Fig.3.

\begin{figure}[ptb]
\begin{center}
\includegraphics[scale=1.1]{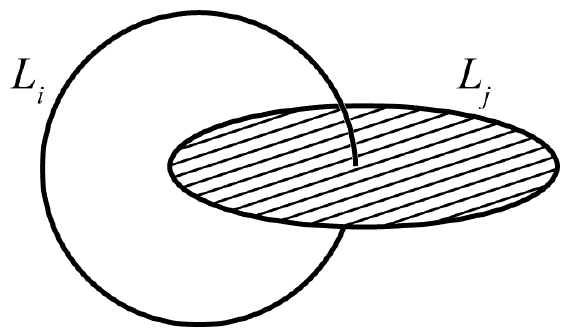}
\end{center}
\caption{Hopf link as intersection of 2-surfaces in 4-space }
\end{figure}

Using this picture, it follows that 
\begin{equation}
S_{K_{1}}\bullet S_{K_{2}}=lk(K_{1},K_{2}).  \tag{3.4}
\end{equation}%
This result is unexpected in the following sense. In standard textbooks on
knot theory$^{41,44}$ we learned that the linking number is a 3-dimensional
object. Now we just found that the linking in 3-dimensions can be
equivalently described in terms of intersecting surfaces "living" in 4
dimensions. This fact allows us to define the \textit{intersection form}$%
^{43}$ Q($\alpha ,\beta )$ for links, say, $\alpha $ and $\beta $ as follows 
\begin{equation}
Q(\alpha ,\beta )=S_{\alpha }\bullet S_{\beta }.  \tag{3.5a}
\end{equation}%
It is an integral matrix since its entries are integers. The matrix $%
Q(\alpha ,\beta )$ is invertible over integers. If $V$ is a (Seifert) matrix
of linking coefficients associated with Seifert surface, then the
intersection form can be represented as ($\func{Re}$f$.56$, page 90) 
\begin{equation}
Q(\alpha ,\beta )=V^{T}-V.  \tag{3.5b}
\end{equation}

The black magic of concordance is revealed now in the following\bigskip\
(absolutely magical)

\textbf{Theorem 3.15. (}Ref.56\textbf{, }page 72\textbf{) }\textit{The
intersection form} (3.5) \textit{is unimodular}, \textit{that is} det $%
Q(\alpha ,\beta )=\pm 1,$\textit{\ if and only if} $\partial $W \textit{is
the disjoint union of integral homology} 3-\textit{spheres}.\bigskip

This result admits simple physical interpretation.\ Indeed, since the
4-manifold is contractible, the linking matrix $Q(\alpha ,\beta )$ should
represent the unlinks/unknots. That is going down to 3 dimensions we have to
deal only with the links/knots living in $\Sigma $ whose mutual linking
numbers are zero. These are exactly the conditions, given without
explanation,\ under which the Casson invariant was defined in$^{55,56,63}$.
Superficially, such a requirement is looking too simple to be physically
interesting. \ This happens \textbf{not} to be the case.\ Below we shall
provide evidence that such an assumption is too simplistic.

At this point we came very close to defining the Casson invariant. To do so
still requires from us to make few steps. This is caused by the fact that
all \ mathematics literature on Casson's invariant, beginning with its first
detailed exposition$^{45}$, defines this invariant purely axiomatically.
Surely, we could do the same here. In such a case the unlinks/unknots
condition as part of the axiomatic package would look to the nonexpert \ as
completely mysterious. \textsl{This circumstance, in part, explains why the
Casson invariant escaped physical interpretation thus far}.

Going back to our discussion, we need to introduce some facts about the
Conway and Alexander polynomials. Although these are obtainable by many
authors in many ways, we prefer to work with the description given in$%
^{44,51}$and$^{64}$. It is based on the presentation of Seifert surfaces for
knots/links as discs with attached bands/ribbons\footnote{%
E.g. see Fig.10}. The mathematical legitimacy of such a presentation of
Seifert surfaces is nicely explained in the book by Massey$^{65}$. In any
case, from these sources we find that: a) for connected sum of knots K$%
_{1}\# $ K$_{2}$ the associated with them Alexander polynomial $\Delta _{%
\text{K}_{1}\#\text{K}_{2}}(t)\dot{=}\Delta _{\text{K}_{1}}(t)\cdot \Delta _{%
\text{K}_{1}}(t)$ . The symbol $\dot{=}$ means "up to a factor" $\pm
t^{n},n\in \mathbf{Z}.";$b) the Alexander polynomial $\Delta _{\text{K}}(t)$
is "blind" with respect to $r$ and $m$ operations defined in section 3.1. \
Mathematically, this blindness is reflected in uses of the mod 2 -type
calculations. Physically, this resembles the difference between the magnets
(where both the direction and the orientation\ with respect to some axis
matters) and the nematic liquid crystals, where only orientation (the angle
with respect to some axis) matters. The classical (Ising-type) spin
naturally admits mod 2 (or Z$_{2})$ description. In the book by Adams$^{66}$%
, on page page 213, \ it is stated \ that " In the case of Ising model, the
resulting partition function yields a knot invariant known as the Arf
invariant." \ The Arf invariant is derivable from both the Alexander and
Conway polynomials and in some references (listed below) it is (mistakenly)
identified with the Casson invariant. Surprisingly, the \ relationship
between the Arf invariant and the Ising model \ is not \ discussed any
further in$^{66}$ . Instead, \ it was established much more recently, in
2012, in$^{67}$. This description is surely not the only way of obtaining
the Arf invariant. \ Many other ways exist. \ For example, it can be also
obtained with help of the Jones polynomial$^{64}$. In this work we shall not
explore this possibility, though, for reasons which will become obvious upon
reading.

Thus we begin with the Conway invariant $\nabla _{\text{K}}$ $(z)$. It is
defined with help of the skein relation$^{64}$ 
\begin{equation}
\nabla _{\text{K}}(z)-\nabla _{\text{\={K}}}(z)=z\nabla _{\text{L}}(z) 
\tag{3.6a}
\end{equation}%
applied to three knots/links K, \={K} and L \ as depicted in Fig.4.

\begin{figure}[ptb]
\begin{center}
\includegraphics[scale=0.9]{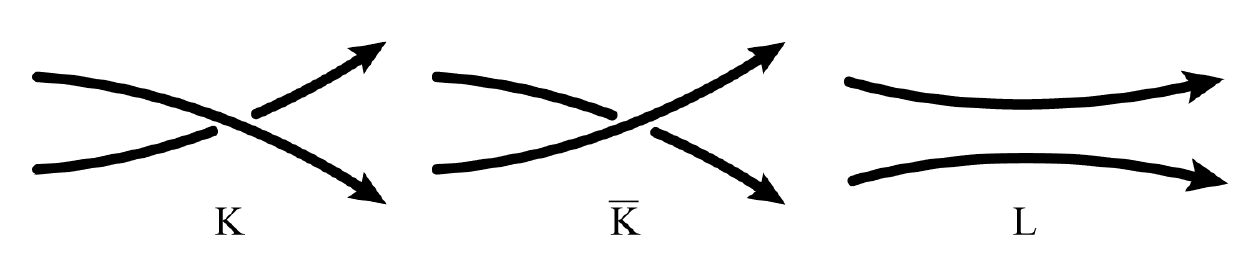}
\end{center}
\caption{Links participating in skein relation}
\end{figure}

By definition, $\nabla _{\text{U}}(z)=1$ (the normalization condition),
where, U is the unknot, as before$.$ The Alexander polynomial $\Delta _{%
\text{K}}(t)$ is obtained from the Conway polynomial by simple replacement z$%
\rightarrow (t^{\frac{1}{2}}-t^{\frac{-1}{2}})$ , e.g. see Ref41, page 110,
resulting in 
\begin{equation}
\Delta _{\text{K}}(t)-\Delta _{\text{\={K}}}(t)=(t^{\frac{1}{2}}-t^{\frac{-1%
}{2}})\Delta _{\text{L}}(t).  \tag{3.6b}
\end{equation}%
The normalization condition is given now$^{64},$page $64,$ for any
(oriented) knot K by $\Delta _{\text{K}}(1)=\pm 1$. Incidentally, this
normalization condition is just a convenience/convention used in knot
theory. In physical applications we are free to choose other normalizations
as long as they are non negative. Furthermore, formally,%
\begin{equation}
\nabla _{\text{K}}(z)=a_{0}(\text{K})+a_{1}(\text{K})z+a_{2}(\text{K}%
)z^{2}+\cdot \cdot \cdot .  \tag{3.7}
\end{equation}%
Here $a_{n}($K$)\in \mathbf{Z,}$ $n=0,1,2,...$ \ Each of $a_{n}($K$)^{\prime
}s$ is (finite-type Vassiliev) invariant$^{68}$ \ of K of order $n$. At the
same time,\ in mod 2-type calculations, each $a_{n}($K$)$ is the concordance
invariant$^{69}$. Following Cohran$^{69}$, we notice that for the $u$%
-component link the expansion (3.7) should be rewritten as 
\begin{equation}
\nabla _{\text{K}}(z)=z^{u-1}(a_{0}(\text{K})+a_{2}(\text{K})z^{2}+\cdot
\cdot \cdot +a_{2n}(\text{K})z^{2n}).  \tag{3.8}
\end{equation}%
Hoste demonstrated$^{70}$ that $a_{0}$ depends only on the pairwise linking
numbers of the components of L while Murakami$^{71}$ demonstrated in accord
with Cohran that, with accuracy up to mod 2 the coefficient $a_{2}$ is the
Arf invariant of L.\ For knots this can be seen by using the analogous
result obtained by Kauffman$^{72}$. He demonstrated that for each knot K ($%
u=1$) the Arf(K)$\equiv a_{2}($K$)$ mod 2. For knots, $a_{0}($K$)=1.$

The Arf(K) invariant can take only two values 0 and 1 dividing all knots
into two classes: those which can be reduced to the unknot U (Arf(U)=0) and
those which can be reduced to the trefoil knot (Arf (K)=1). More on this
will be said in the next section. \ Just like the signature $\sigma ($K$),$
the Arf(K) is invariant of concordance since it vanishes on slice knots.
This means that it behaves like the signature with respect to the operation
of taking the connected sum (3.3). While the signature $\sigma ($K$)$
provided us with the mapping (a homomorphism) $\sigma ($K$)\rightarrow 
\mathbf{Z}$ (or C$_{3}\rightarrow \mathbf{Z)},$the Arf(K)$\in \mathbf{Z}%
_{2}=\{0,1\}$ is providing us with a homomorphism $C_{3}\rightarrow \mathbf{Z%
}_{2}.$

The information just described is sufficient formally for introduction of
the Casson invariant. It is given by the following \bigskip

\textbf{Theorem 3.16}.(Ref.$50$, page 88) \textit{The Casson invariant
coincides with the second coefficient of the Conway polynomial} (i.e. with $%
a_{2}($K$)).$ \hspace{0.1in}\bigskip

\textbf{Remark 3.17. }There is no mention of mod 2 in this theorem. In
addition, this \ definition of the Casson invariant formally is not in
accord with those given in$^{45,55.56.63,73}$ . Last but not the least,
taken mod 2 the coefficient $a_{2}($K$)$ is coinciding with the Arf(K)
invariant which is a concordance invariant while we mentioned already that
the Casson invariant is not the concordance invariant, e.g. read Ref.45 page
XV (bottom).\medskip

Thus, now we have to discuss other definitions of Casson's invariant and how
they are related to that given in$^{50}$ . This \ discussion will be
helpful\ for us in a number of ways.

We begin with the following observations. \ Let $V$ be $2n\times 2n$ Seifert
matrix for some knot/link K in some basis with $n=2g+u-1$, where $g$ is the
genus of Seifert's surface. Using $V$ the Alexander polynomial $\Delta _{%
\text{K}}(t)$ can be defined as$^{74}$, page 112, 
\begin{equation}
\Delta _{\text{K}}(t)=\det (V^{T}-tV).  \tag{3.9}
\end{equation}%
Following Kauffman$^{44}$, page 200, it is convenient to introduce the
potential function of K as 
\begin{equation}
\Omega _{\text{K}}(t)=\det (t^{-1}V-tV^{T})=\pm t^{2n}\det
(V^{T}-t^{2}V)=\pm t^{2n}\Delta _{\text{K}}(t^{2}).  \tag{3.10}
\end{equation}%
When applied to three knots/links K, \={K} and L depicted in Fig.4, the
skein relation for $\Omega _{K}(t)$ is given by 
\begin{equation}
\Omega _{\text{K}}(t)-\Omega _{\text{\={K}}}(t)=(t-t^{-1})\Omega _{\text{L}%
}(t)  \tag{3.11}
\end{equation}%
Comparison between eq.s (3.6a) and (3.11) yields 
\begin{equation}
\Omega _{\text{K}}(t)=\nabla _{\text{K}}(t-t^{-1}).  \tag{3.12}
\end{equation}%
At the same time, we already have an equation (3.6b). Therefore, 
\begin{equation}
\Delta _{\text{K}}(t)=\Omega _{\text{K}}(\sqrt{t})  \tag{3.13}
\end{equation}%
By comparing (3.12) and (3.13) we obtain, 
\begin{equation}
\Delta _{\text{K}}(t)\dot{=}\nabla _{\text{K}}(\sqrt{t}-\sqrt{t^{-1}}). 
\tag{3.14}
\end{equation}%
In addition, \ it is important to notice that $\Delta _{\text{K}}(t)\dot{=}%
\Delta _{\text{K}}(t^{-1}).$ Indeed,%
\begin{equation*}
\Delta _{\text{K}}(t)=\det (V^{T}-tV)=\left( -t\right) ^{2n}\det
(-t^{-1}V^{T}+V)\dot{=}\det (-t^{-1}V^{T}+V)^{T}=\det (V^{T}-t^{-1}V)
\end{equation*}%
From this property it follows that 
\begin{eqnarray*}
\Delta _{\text{K}}(t) &=&b_{-n}(\text{K})t^{-n}+b_{-(n-1)}(\text{K}%
)t^{-(n-1)}+\cdot \cdot \cdot +b_{(n-1)}(\text{K})t^{(n-1)}+b_{n}(\text{K}%
)t^{n}\text{ \ } \\
\text{implying }b_{-n}(\text{K}) &=&b_{n}(\text{K}),b_{-(n-1)}(\text{K}%
)=b_{-(n-1)}(\text{K}),...,b_{-1}=b_{1}.
\end{eqnarray*}%
Therefore, using this result the canonical form for the Alexander polynomial
is known to be$^{74}$ 
\begin{equation}
\Delta _{\text{K}}(t)=\tilde{a}_{0}(\text{K})+\tilde{a}_{1}(\text{K}%
)(t+t^{-1})+\cdot \cdot \cdot +\tilde{a}_{n}(\text{K})(t^{n}+t^{-n}). 
\tag{3.15a}
\end{equation}%
In view of equations (3.6b) and (3.8), in the case of knots we formally have
instead 
\begin{eqnarray}
\Delta _{\text{K}}(t) &=&\tilde{a}_{0}(\text{K})+\tilde{a}_{2}(\text{K})(%
\sqrt{t}-\sqrt{t^{-1}})^{2}+\cdot \cdot \cdot +\tilde{a}_{2n}(\text{K})(%
\sqrt{t}-\sqrt{t^{-1}})^{2n}  \TCItag{3.15b} \\
&\equiv &\mathfrak{b}_{0}(\text{K})+\mathfrak{b}_{1}(\text{K}%
)(t+t^{-1})+\cdot \cdot \cdot +\mathfrak{b}_{n}(\text{K})(t^{n}+t^{-n}). 
\TCItag{3.15c}
\end{eqnarray}%
Now we are in the position to refine Theorem 3.16 and to introduce the
Casson invariant correctly. For this purpose we have to rewrite the
expansion for the Alexander polynomial in (3.15b) in the form used for the
Conway polynomial, e.g. (3.8) (with $u=1$). That this is possible can be
seen from general results presented in$^{74}$, page 113. Since this is
possible, we formally \ obtain:%
\begin{equation}
a_{2}(\text{K})=\frac{1}{2}\frac{d^{2}}{dz^{2}}\Delta _{\text{K}}(z)\mid
_{z=0}\equiv \frac{1}{2}\Delta _{\text{K}}^{\prime \prime }(1),  \tag{3.16}
\end{equation}%
because $z=0$ is the same as $t=1$. Thus, we need to solve the following
problem: given \ the Conway-like expansion for the Alexander polynomial for
the knot K,%
\begin{equation}
\Delta _{\text{K}}(z)=a_{0}(\text{K})+a_{2}(\text{K})z^{2}+\cdot \cdot \cdot
+a_{2n}(\text{K})z^{2n},  \tag{3.17}
\end{equation}%
find coefficients $\mathit{a}_{0}($K$),a_{2}($K$),...$in $(3.17)$ in terms
of the coefficients $\mathfrak{b}_{0}($K$),\mathfrak{b}_{1}($K$),\mathfrak{b}%
_{2}($K$),...,\mathfrak{b}_{n}($K$)$ in (3.15c). This was almost done in$%
^{44}$, pages 205-206. However, Ref.44 contains some misprints causing us to
redo calculations. Clearly, we are only interested in calculating $\mathfrak{%
b}_{1}($K$).$ In view of equations (3.11)-(3.14) we introduce the following
identification: $z=\sqrt{t}-\sqrt{t^{-1}}$. With help of such an
identification \ we obtain: 
\begin{eqnarray*}
t+t^{-1} &=&z^{2}+2, \\
t^{2}+t^{-2} &=&z^{4}+4z^{2}+2, \\
t^{3}+t^{-3} &=&z^{6}+6z^{4}+9z^{2}+2, \\
&&.... \\
t^{n}+t^{-n} &=&z^{2n}+2nz^{2n-2}+\cdot \cdot \cdot +2.
\end{eqnarray*}%
With help of these results we obtain: 
\begin{equation}
\mathit{a}_{2}(\text{K})=\mathfrak{b}_{1}(\text{K})+4\mathfrak{b}_{2}(\text{K%
})+9\mathfrak{b}_{3}(\text{K})+\cdot \cdot \cdot +n^{2}\mathfrak{b}_{n}(%
\text{K})\equiv \dsum\limits_{i=1}^{n}i^{2}\mathfrak{b}_{i}(K).  \tag{3.18}
\end{equation}%
We would like to test just obtained results using \ known results for the
trefoil knot. In this case the Conway polynomial is given by $\nabla _{\text{%
K}}(z)=1+z^{2}$ \ while the Alexander polynomial is given by $\Delta _{\text{%
K}}(t)=1+(\sqrt{t}-\sqrt{t^{-1}})^{2}=t-1+t^{-1}$ . Accordingly, $a_{2}($K$%
)=1,$ in view of the definition (3.16). \ Thus, we just reobtained well
known standard results for the trefoil$^{44,56,64}$ .

Now we are ready to define the Casson invariant of homology spheres. Let M
be an oriented homology 3-sphere (e.g. complement of the standard slice knot
in $S^{3}).$ \ It is described by the \textit{integer-valued} invariant $%
\lambda ($M$)-$the Casson invariant. Notice that for slice knots/links $%
\lambda ($M$)\neq 0.$ Therefore it is not a concordance invariant (in accord
with page XV of Ref.45.). It possesses the following properties:

1) For $S^{3},$ that \ is when $\pi _{1}($M$)=1,$ $\lambda (S^{3})=0.$
Incidentally, $\pi _{1}($M$($K$,\alpha ))=1$ only when $\alpha =0$ (e.g.
read Remark 3.9 and Ref.59, page 154).

2) $\lambda (-$M$)=-\lambda ($M$).$

3) If K$_{2}$ is knot in M (e.g. read Theorem 3.13 (b)) and K$_{1}$ is slice
knot, then M(K$_{1}$,K$_{2}$;$\alpha ,\beta .\gamma ,\delta $) is homology
sphere \ if $\left\vert \gamma \right\vert =1.$ Since det A=-1 we obtain $%
\alpha \delta -\beta \gamma =\alpha \delta -\pm \beta =-1$

that is $\beta =\pm (\alpha \delta +1).$ Therefore, we can can choose $%
\alpha $ and $\delta $ as independent variables along with the sign $%
\epsilon =\pm 1.$ This allows us to relabel M(K$_{1}$,K$_{2}$;$\alpha ,\beta
,\gamma ,\delta $) as M$^{\epsilon }$(K$_{1}$,K$_{2}$;$\alpha ,\delta $).

Following$^{45,59}$ consider a special case M(K$_{1}$,K$_{2}$;$\alpha ,\beta
.\gamma ,\delta $)=M(K$_{1}$,K$_{2}$;$-1,0,\pm 1,1$) and introduce the
notation%
\begin{equation}
M(\text{K}_{1},\text{K}_{2};-1,0,\pm 1,1)\simeq ((\text{K}_{1}\#\text{K}%
_{2})_{\pm 1};\text{M}_{1}\#\text{M}_{2})\equiv (\text{\={K}}_{\pm 1};\text{%
\={M}}).  \tag{3.19}
\end{equation}

\ \ Taking into account the Remark 3.9, consider a knot \={K} in a homology
sphere \={M} so that the symbol (\={K}$_{n}$;\={M}) \ will denote an
oriented homology sphere obtained by performing $\pm /n$ Dehn surgery on \={M%
} along \={K}, $n\in Z.$ Then, the difference (the derivative) is defined as 
\begin{equation}
\lambda ^{\prime }(\text{\={K}},\text{\={M}})=\lambda (\text{\={K}}_{n+1};%
\text{\={M}})-\lambda (\text{\={K}}_{n};\text{\={M}}).  \tag{3.20}
\end{equation}

\ \ \ \ Because thus defined discrete "derivative" is n-independent (this
crucial property will be used essentially below), by applying the induction
\ we obtain:

4) \ 
\begin{equation}
\lambda (\text{\={K}}_{n};\text{\={M}})=\lambda (\text{\={M}})+n\lambda
^{\prime }(\text{\={K}},\text{\={M}).}  \tag{3.21}
\end{equation}

\ \ implying, $\lambda ($\={K}$_{0};$\={M}$)=\lambda ($\={M}$).$

\ \ \ If in addition we require that

5) 
\begin{equation}
\lambda (\text{M}_{1}\#\text{M}_{2})=\lambda (\text{M}_{1})+\lambda (\text{M}%
_{2}),  \tag{3.22}
\end{equation}

\ \ then in view of (3.19) we obtain:

\begin{equation}
\lambda (((\text{K}_{1}\#\text{K}_{2})_{n};\text{M}_{1}\#\text{M}%
_{2}))=\lambda ((\text{K}_{1}\text{)}_{n};\text{\={M}}_{1})+\lambda ((\text{K%
}_{2}\text{)}_{n};\text{\={M}}_{2}).  \tag{3.23}
\end{equation}

\ \ By using equations (3.19)-(3.22) we obtain $\lambda ($M$^{\epsilon }($K$%
_{1},$K$_{2};-1,1))=\lambda ($M$_{1})+\epsilon \lambda ^{\prime }($K$_{1};$M$%
_{1})+\lambda ($M$_{2})+\epsilon \lambda ^{\prime }($K$_{2};$M$_{2}).$

To connect the obtained results with the content of Theorem 3.16 the
following observations are helpful. Following Ref.44 we notice that for
coefficients of the Conway polynomial defined by equation (3.7) the
following recursion formula takes place

\begin{equation}
a_{n+1}(\text{K})-a_{n+1}(\text{\={K}})=a_{n}(\text{L})  \tag{3.24}
\end{equation}%
where, as before, K, \={K} and L are the same as depicted in Fig.4. \
Therefore 
\begin{equation}
a_{0}(\text{K})=\left\{ 
\begin{array}{c}
1\text{ \ \ \ \ \ \ \ \ \ \ \ \ \ \ if K has one component} \\ 
0\text{ \ if K has more than 1 component }%
\end{array}%
\right.  \tag{3.25 }
\end{equation}%
and 
\begin{equation}
a_{1}(\text{K})=\left\{ 
\begin{array}{c}
lk(\text{K})\text{ \ \ \ \ \ \ \ \ \ \ \ \ \ \ \ if K has two components} \\ 
0\text{ \ \ \ \ \ \ \ \ \ \ \ \ \ \ \ \ \ \ \ \ \ \ \ \ \ \ \ \ \ \ \ \ \ \
\ \ \ \ \ \ \ \ otherwise}%
\end{array}%
\right.  \tag{3.26}
\end{equation}%
In the simplest case we obtain, 
\begin{equation}
a_{2}(\text{K})-a_{2}(\text{\={K}})=lk(L).  \tag{3.27}
\end{equation}%
Equivalently, just obtained results can be rewritten in terms of the second
derivatives of Alexander polynomial with help of equation (3.16). Since $%
a_{2}($K$)$ mod 2 is Arf (K) invariant and we already noticed that Arf (K)
is the concordance invariant, this implies that in $S^{3}$ we should have%
\begin{equation}
a_{2}(\text{K}_{1}\#\text{K}_{2})=a_{2}(\text{K}_{1})+a_{2}(\text{K}_{2}). 
\tag{3.28}
\end{equation}%
Now it remains to compare (3.22) with this result. \ This causes us \ to
check/to prove whether or not

\begin{equation}
\lambda ^{\prime }(\text{K}_{1}\#\text{K}_{2};\text{M}_{1}\#\text{M}%
_{2})=\lambda ^{\prime }(\text{K}_{1};\text{M}_{1})+\lambda ^{\prime }(\text{%
K}_{2};\text{M}_{2}).  \tag{3.29}
\end{equation}%
If this, indeed, can be proven, \ then using equations (3.16),(3.28) and
(3.29) it is possible to make an identification.

\begin{equation}
\lambda ^{\prime }(\text{K};\text{M})=\frac{1}{2}\Delta _{\text{K},\text{M}%
}^{\prime \prime }(1).  \tag{3.30}
\end{equation}%
This result was obtained for the first time by Casson$^{45,56}$.
Equivalently, using definition of the discrete derivative (3.20) and taking
into account that such a derivative is n-independent, equation (3.30) can be
equivalently rewritten in the form 
\begin{equation}
\lambda (\text{K}_{n};\text{M})=\lambda (\text{M})+\frac{n}{2}\Delta _{\text{%
K},\text{M}}^{\prime \prime }(1).  \tag{3.31}
\end{equation}%
This relation (not an equation!)\ admits physical interpretation to be
discussed in the next subsection.\bigskip \medskip

3.3. \ \ Regge mass spectrum from the Casson invariant (a hint)

\bigskip

In our previous work$^{15}$ we took advantage of the fact that the path
integral for pure Y-M gauge fields, when treated nonperturbatively, is
reducible to the topological C-S field theory. By applying the Abelian
reduction it\ is converted into the hydrodynamics/ ideal
magnetohydrodynamics -type model functional whose detailed study is
presented in the companion work$^{22}$. Connections of the C-S model with
the Kontsevich invariants/integrals and the Vassiliev-type invariants,
including $a_{2}($K$),$ is discussed in detail in$^{50,75}$. For reasons
explained in the previous subsection, we cannot use these results.
Furthermore, we \ also cannot use these results because treatments of the
Casson invariant presented in$^{50,75}$are intrinsically perturbative. They
are distinctly different from the instanton (Floer)-type nonperturbative
treatments presented in$^{16,17}$. We would like to remind to our readers
that the basics of instanton (Floer) approach, as explained in our work,$%
^{15}$ is done in full compliance with much more comprehensive treatments$%
^{13,14}$. In view of the (Mirror) Theorem 3.3. uses of istantons in the
present case are as essential as their uses in analogous situations, e.g. in
chemical reactions$^{76}$. In this work \ we only present needed arguments
for such instanton-type treatment. It will be developed in the future
publications. In this work the results are presented at the rigorous
mathematical level inspired by known phenomenological results. \ 

We begin the description of these results with a gentle reminder of the
results of nonrelativistic scattering theory. In it, the scattering
amplitude is expected to possess the (Regge-type) poles in the complex
angular momentum $J$ plane. The pole equation $J=\alpha (E)$ relates $J$ to
the energy $E.$ The function $\alpha (E)$ is called \textsl{the Regge
trajectory. }In the relativistic case, the energy parameter $E$ is replaced
by the appropriate Mandelstam variable, say, $s$. Ref.77 is describing the
basic steps for building the dual resonance models\ using the Regge
phenomenology as an input. The first push \ towards development of \ dual
resonance models was given by Veneziano$^{78}$ in 1968. In that year he
postulated that the meson-meson scattering amplitudes can be described in
terms of combination of Euler's Beta functions symmetric with respect to
permutation of its\ arguments. These are the Mandelstam variables $s,t,u$.
Search for a model reproducing these amplitudes resulted in all kinds of
string models as is well known. The output of these relativistic models is
scattering amplitudes possessing the Regge-type poles. In connection with
this result several questions emerge. \ The \ first among them is this: How
string models are related to quantum chromodynamics (QCD)? That is to say,
can string models be derived from the QCD or, vice-versa, can QCD be derived
from the sting models? \ Since this is not a review paper, we mention only
the very latest works relevant to this question. In particular, we begin with%
$^{79,80}$. In Ref.79 the authors consider results originating from the
diquark model. Quoting them, "At a crude level, the idea is that pairs of
quarks form bound states which can be treated as (confined) particles....It
is plausible that baryons with large values of the angular momentum $J$ form
extended bar-like structures, with quarks pushed to the extremities by
centrifugal forces". The bar-like \ structures are made of electric flux
tubes, or strings. The results of$^{79}$ were carefully analyzed in$^{80}$.
It happens that the comparison between theory and experiment (including
computer simulations) \ can be made by using the Chew-Frautschi (C-F) formula%
\begin{equation}
M^{2}=a+bJ.  \tag{3.32}
\end{equation}%
Here $a$ and $b$ are some constants, $M$ is the hadron (baryon or meson)
mass and the angular momentum $J=n$, $n=0,1,2,...$ The above formula is
basically the same thing as the Regge trajectory: $\alpha (t)=\alpha
(0)+\alpha ^{\prime }t,$ $\alpha ^{\prime }>0.$ Results of Ref.80 indicate
that the C-F formula compares exceptionally well with experimental data for
both baryons and mesons. From the theoretical side, the interest lies in
obtaining the values of parameters $a$ and $b$. As demonstrated in Ref.80,
in the limit of zero quark masses, the diquark model$^{79}$ compares very
well with the C-F formula and, therefore, with the experiment. The results
obtained in$^{79}$ were reproduced and further improved in$^{81,82}$ with
the purpose of obtaining the values of $a$ and $b$ theoretically. Contary to
the initial expectations, experimental data \ convincingly suggest that the
slope parameter $b$ is the same for both baryons and mesons, \textsl{even if
the quark masses are not vanishingly small!} The slope was calculated in$%
^{81,82}$ and was fitted to the C-F formula\ with reasonable degree of
success. The value of intercept $a$ was calculated in$^{83}$, where entirely
different model (the conventional bosonic string model) was used and,
accordingly,\ the entirely different calculations (more traditional,
string-theoretic) were done, in dimensions 4 and higher. The obtained
results are much less satisfactory though. Notice, the obtained results do
not provide an answer to the question posed above. Moreover, the results of
sections 1 and 2 are also not helpful. The second question is the following:
\ To what extent the C-F formula confirms or denies the existence of
instantons in QCD? \ 

To our knowledge, at the moment it is possible, in principle, to give
answers both ways. For instance, if one believes that strings models can
incorporate gravity (not renormalizable by the conventional methods), then
the QCD should be derivable from the string models. However, the
abelianization procedures discussed in section 2 indicate that \textsl{all}
gauge fields: electromagnetic, Yang-Mills and gravity, admit the same type
of treatment, e.g. outlined in our works$^{15,22,84}$. And, if this \textsl{%
is} the case, then the QCD is reduced to the C-S topological field theory
and its Abelian version is apparently sufficient. That is to say, in such a
case the problem of recovering of string model from QCD was to a large
extent solved already by Nambu$^{85}$. In our paper$^{86}$ some practical
applications (other than in high energy physics) of this line of thought
were discussed in detail. It is being hoped, that physical insight coming
from fields other than high energy physics could be helpful for resolving
problems of high energy physics.

Notice also that dynamically generated \textsl{nonhyperbolic} knots and
links are observables in the C-S field theory. In fact, without asking the
question about how these knots/links were created, for the C-S model all
knots and links are observables$^{87}$. The fermionic effects can be
modelled in such an environment purely topologically$^{88,89}$. \ Since the
black holes can be described in terms of elementary particles$^{90},$it
appears that no fields other than gauge fields (Abelian or not) are needed.
The effects of charges can be included consistently into such formalism$%
^{15,18,22}$.

Now we are in the position to compare the C-F formula (3.32) with the 
\textsl{Casson surgery formula} (3.31) \footnote{%
Notice, in mathematics literature \ the equation (3.31) is known as the 
\textsl{\ Casson surgery formula}.}. Clearly, they formally coincide. The
question remains: What is the physical content of the Casson Surgery
formula? The answer is provided in the rest of this paper.\bigskip

\bigskip

{\Large 4. \ }{\large Physics behind the Casson surgery formula}

\bigskip \bigskip

4.1. \ Physical content of the Dehn surgery.

\ \ \ \ \ \ \ \ From Moffatt intuition to Hempel, Rolfsen and Kirby proofs

\bigskip

The Casson surgery formula (3.31) involves the notion of Dehn surgery, e.g.
the symbol (\={K}$_{n}$;\={M}) \ defined in section 3 denotes an oriented
homology sphere obtained by performing $\pm /n$ Dehn surgery on \={M} along 
\={K}, $n\in Z.$ Based on this information, the question emerges: Since the
C-F and Casson surgery formula look identical, could this be just a
coincidence or, could it be that the Casson surgery formula indeed carries
some hidden physics in it? We would like to demonstrate that, indeed, this
is the case. For this we have to demonstrate that the elementary Dehn
surgery can be looked upon (equivalent to) as an operation of crossing
change depicted in Fig.5.\medskip

\begin{figure}[ptb]
\begin{center}
\includegraphics[scale=1.4]{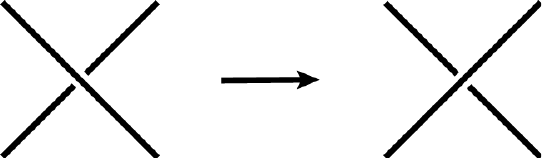}
\end{center}
\caption{}
\end{figure}

From this standpoint the basic skein relation for the Vassiliev invariants,
Fig.2, can be physically interpreted as follows. Imagine a knot $K$ made of
rope, a flux tube, say. Since in this work we relate knots to particles,
different knots/links are obtained from each other by cutting the rope and \
regluing it back. Each time such an operation is performed, \ it is
associated with the fixed once and for all amount of energy. Thus the basic
skein relation for the Vassiliev invariants is graphical illustration of
this process. In the case of Casson surgery formula, the analog of such
skein relation is given by the combined use of equations (3.20) and (3.30),
that is by 
\begin{equation}
\lambda (\text{K}_{n+1};\text{M})-\lambda (\text{K}_{n};\text{M})=\frac{1}{2}%
\Delta _{\text{K},\text{M}}^{\prime \prime }(1),  \tag{4.1}
\end{equation}%
provided that we can prove that the elementary Dehn surgery can be replaced
\ by \ the elementary crossing change. We shall do just this now.

We begin with some excerpts from the paper by Livingston$^{91}$. Recall that
in the Conway polynomial expansion (3.7) each of a$_{n}$(K)'s is \
finite-type Vassiliev invariant of K of order n. Take into account equations
(3.24), (3.25) and (3.26). Then, the skein relation depicted in Fig.2
acquires the following look:%
\begin{equation}
lk(K)-lk(\bar{K})=1.  \tag{4.2}
\end{equation}%
This result is illustrated in Fig.6 on example of the Hopf link.\ 

\begin{figure}[ptb]
\begin{center}
\includegraphics[scale=1.0]{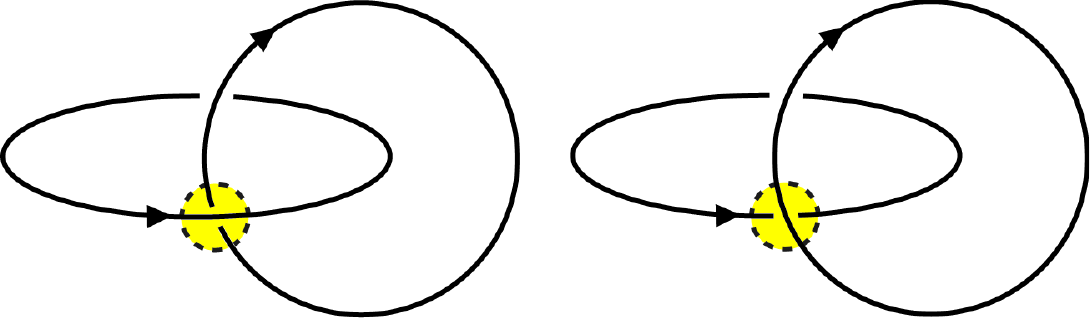}
\end{center}
\caption{}
\end{figure}

Notice that in this case \={K} is unlinked and, therefore, $lk(\bar{K})=0$.
Now, take into account equations (3.17) and (3.27) and apply them to the
case of the trefoil knot K as depicted in Fig.7\ 

\begin{figure}[ptb]
\begin{center}
\includegraphics[scale=1.0]{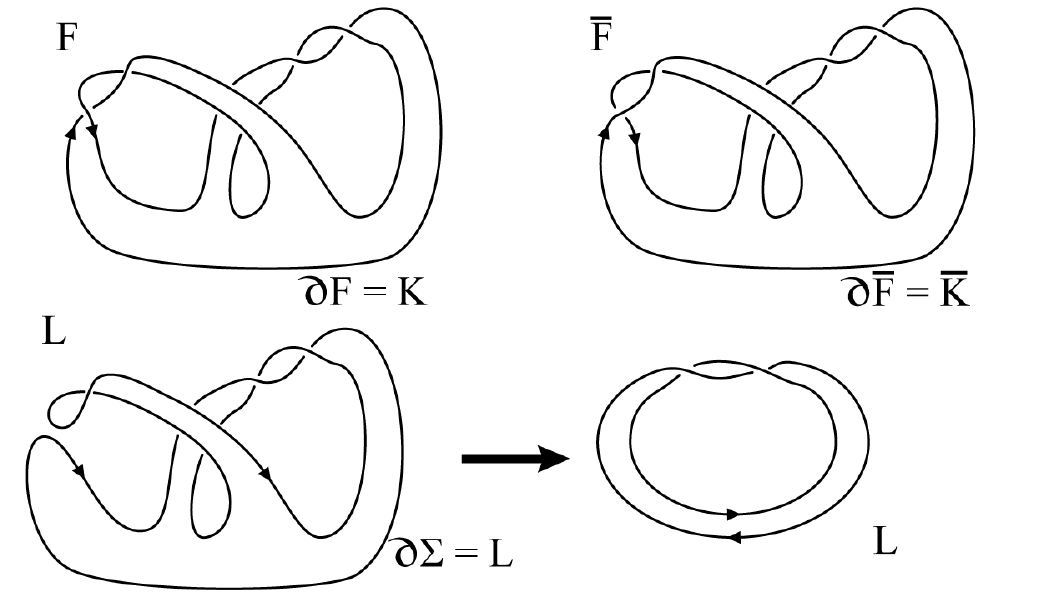}
\end{center}
\caption{ Illustration of the skein relation (3.27) for the Seifert surface
F of the trefoil knot K}
\end{figure}

In such a case we obtain:%
\begin{equation}
a_{2}(\text{K})=lk(\text{L)}  \tag{4.3}
\end{equation}%
since $a_{2}($\={K}$)=0$ \ (because \ \={K} is unknot) and L is the Hopf
link. Using this result in (4.1) while taking into account (3.16), (3.27)
and (3.30) we just obtained a very plausible result strongly hinting at the
connection between the operation of elementary crossing change and of the
elementary Dehn surgery. These suggestive observations we would like now to
convert into a solid proof. In view of its fundamental importance for this
work, we would like to put the content of this proof into physical frame
following ideas of Moffatt$^{92}$. In this brief note he said the following:
"Any knot or link may be characterized by an `energy spectrum'- a set of
positive real numbers determined solely by its topology. The lowest energy
provides a possible measure of knot or link complexity." The energy spectrum
is obtained by the operation which Moffatt describes as follows. 1. Surround
knot by a tubular neighborhood (e.g. read the Appendix ), that is by the
solid torus. The "magnetic" (Abelian or non Abelian) flux circulating inside
this torus is carrying some energy (e.g. read$^{15}$). 2. The flux tube is
cut at any section ' $\varphi =const$. 3.Thus formed cylinder is twisted
through the angle $2\pi h_{0}$. 4. After this, the tube is reconnected. In
Figure 2 of$^{92}$ \ Moffatt provides a physical sketch of how this process
can actually take place in real world. He observes that "The unknotted tube $%
\mathcal{T}_{0}$ is converted into the knotted tube $\mathcal{T}_{K}$ by
switching a number of crossings." This knotting/unknotting is facilitated by
the unknots which appear from nowhere and disappear into nowhere in
Moffatt's paper. He writes : " Each switch is equivalent to the insertion of
a small loop that cancels the field on one side\footnote{%
Of the flux tube (our comment)} and makes it to reappear on the other. The
change in helicity $h$ associated with the switches is $..h=h_{0}-4$." \
From Arnol'd \ inequality, equation (3.15) of$^{15}$, it follows that the
flux energy is bounded from below by the helicity with equality achieved for
the force-free \ fields$^{15,22}$. This means that sequential crossing
changes produce the energy spectrum $h=h_{0}\pm 4n$, $n=0,1,2,...$ Clearly,
this result is the C-F formula (3.32).

Being armed with these physically motivating arguments, we are ready to
bring them into correspondence with rigorous mathematics. The major issue
requiring clarification is associated with the insertion /deletion of small
loops "that cancel the field on one side and make it to reappear on the
other". Our readers are encouraged to read the Appendix at this point. With
this assumption, the operation of a single crossing change is depicted in
Fig.8.

\begin{figure}[ptb]
\begin{center}
\includegraphics[scale=1.0]{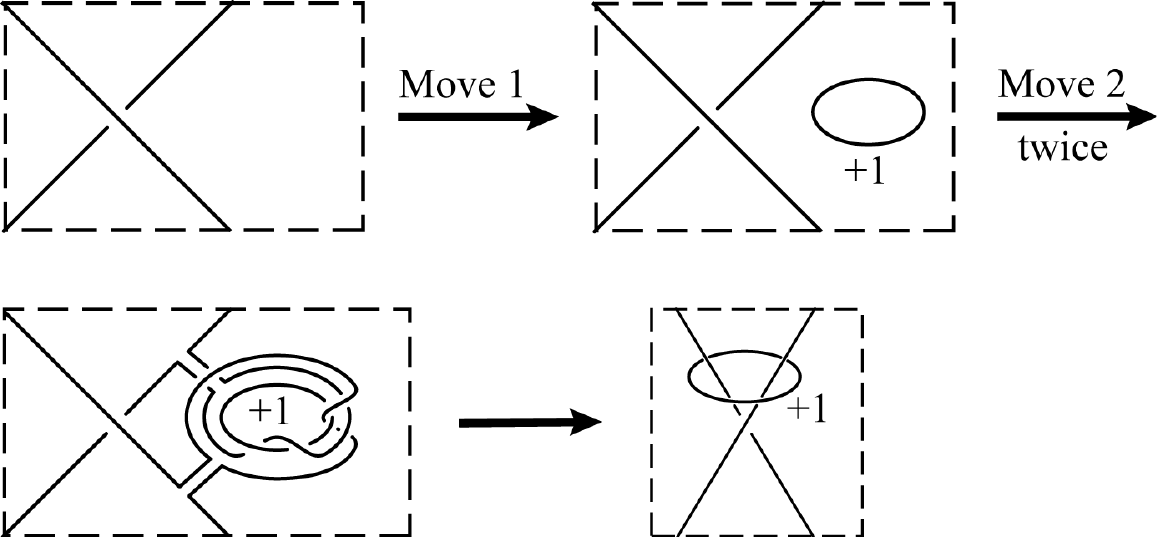}
\end{center}
\caption{ Change of crossing induced by the 1st (Move 1) and the 2nd (Move
2) Kirby moves}
\end{figure}

It requires uses of both types of Kirby moves as depicted. The same result
was obtained by Hempel$^{93}$ in 1961 by different arguments. This result
was published long before the Kirby calculus were invented by Kirby$^{94}$
in 1978. Hempel's result was used by Rolfsen$^{95}$ in 1974 for development
of far reaching surgical interpretation of the Alexander polynomial. The
idea of Hempel's proof of equivalence of elementary Dehn surgery to crossing
change is depicted in Fig.9.

\begin{figure}[ptb]
\begin{center}
\includegraphics[scale=1.0]{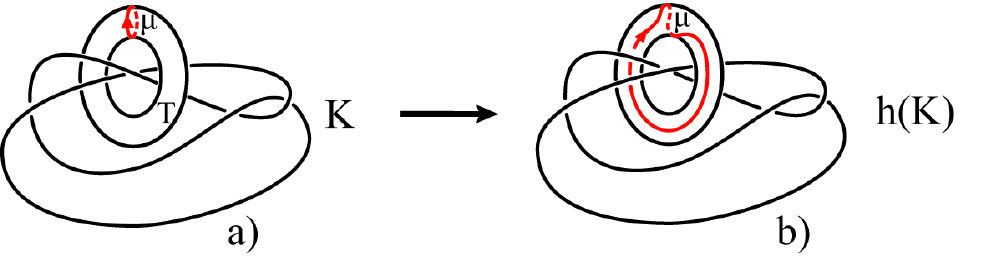}
\end{center}
\caption{ Change of crossing induced by the elementary Dehn twist performed
on the torus associated with the unknot}
\end{figure}

4.2. Some important ramifications\bigskip

The results of previous subsection should be used with some caution. This is
so because of several reasons. Our readers should not confuse the physical
picture suggested by Moffatt with the actual picture in the context of
Casson surgery formula. If one follows Moffatt's ideas, then one should
start with some knot K and apply to it successfully the crossing change
operations. If one begins with the unknot U, then the tower of excited
states is obtained by producing knots of ever increasing complexity. If,
instead, one begins with some nontrivial knot then this knot relaxes to the
unknot via crossing changes mechanism. In the Casson case physics is
different. From the description already presented it follows that instead of
crossing changes one should use the sequence of Dehn twists. To our
knowledge, there is no direct physical analog of the Dehn twists, even if
one is using some viscous elastic medium. However, the results depicted in
Fig.s 8,9 indicate that it is perfectly legitimate to identify the
elementary Dehn twist with the crossing change operation. This type of
homeomorphism admits physical interpretation analogous to that developed by
Moffatt. Analogous but not identical! And, therefore, this new
interpretation is much more suitable for the high energy physics
applications. If one follows Moffatt's ideas, then every knot/link relaxes
to its ground state -the unknot. In the present case, the Dehn surgery
operation on the unknot in $S^{3}$ results in unknot. By performing the
elementary Dehn surgery on the simplest nontrivial knot-trefoil, one is
obtaining the dodecahedral Poincare$^{\prime }$ space- a homology sphere.
This means that relaxation process-from some homology sphere-back to the
space in which the trefoil lives will end up in a stable particle associated
with the trefoil. This makes perfect sense physically. Indeed, both, in our
work$^{22}$ and in$^{2}$ the trefoil emerges as the stable ground state
particle. In addition, recent attempt to build the standard model based on
different labelings of the trefoil knot was developed in$^{96}$ (and
references therein). Between$^{2,22}$ and$^{96}$, only$^{22}$ uses rigorous
mathematical results consistent with results by Floer and Taubes$^{16}.$
Only this development is allowing us to bring ultimately into play the
concept of concordance. The result for $\Delta _{\text{K},\text{M}}^{\prime
\prime }(1)$ entering (3.31) was obtained with help of the Alexander
polynomial calculated for a knot K in 3-manifold M. How such a polynomial
can be calculated in M if all textbooks on knot theory provide us only with
calculations of this polynomial for $\ $knots$\ $\ in $S^{3}$? \ Suppose we
calculate $\Delta _{\text{K, S}^{3}}(t)$. Can this information be used for
calculation of $\Delta _{\text{K,M}}(t)$? Could it be that $\Delta _{\text{%
K,M}}(t)=$ $\Delta _{\text{K, S}^{3}}(t)$ ?

To prove or disprove this equality we have to recall the protocol for
designing of the Alexander polynomial. It begins as follows. 1.We associate
with a given knot/link the Seifert surface. 2. This surface always can be
presented as disc with ribbons, e.g. see Fig.10.\ 

\begin{figure}[tbp]
\begin{center}
\includegraphics[scale=1.0]{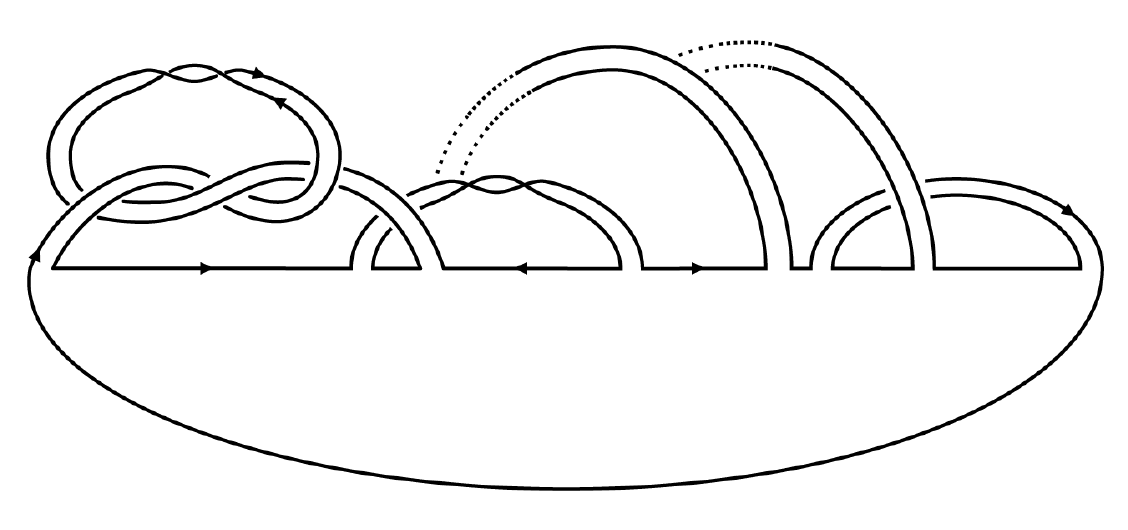}
\end{center}
\caption{ A typical Seifert surface for some knot or link}
\end{figure}

3. By known rules, using such a disc with ribbons in a standard way
described in any knot theory textbook it is possible to obtain the Seifert
matrix $V$ (e.g. see section 3) and, subsequently, with its help-the
Alexander polynomial, e.g. see (3.9), or its equivalent, the potential
function (3.10). 4. The ribbons attached to the disc could be knotted and
linked with each other, and each of them could be twisted an even number of
times (since the Seifert surface is orientable).Topologically, the effects
of Kirby moves (e.g. read Appendix) exhibit themselves via sliding one of
the attachment points of a given ribbon over another ribbon and reconnecting
it back to the disc without changing the isotopy of its boundary. The
resulting surface is again a disc with ribbons. This sliding move is
understood the best in the context of the theory of 4-manifolds$^{43}$, or,
alternatively, Ref.97, Chapter 5. For homology sphere 3-manifolds obtained
via $\pm $1 surgery on framed links Habiro$^{98}$elegantly refined Kirby
calculus using theory of 4-manifolds.To such modified disc with ribbons it
is permissible to add two additional bands. One of them is untwisted and
unknotted while the other could be twisted and/or knotted and can link
another ribbons as schematically depicted in Fig.10. Topologically, however,
for the links such an addition does not affect the boundary of the Seifert
surface (this boundary is made out of the link in question) so that the
Seifert surface with two extra ribbons still represents the same link. All
this is beautifully explained in the Kauffman's book$^{44}$, pages 196,197.
The operation of adding or subtracting these two bands is called \textsl{%
stabilization}. Two Seifert surfaces are \textsl{stably equivalent} if there
is a sequence of stabilizations applied to each of them so that they can be
deformed into each other. It is known$^{56,64}$ that any compact orientable
3-manifold M can be obtained via surgery on a link $\mathcal{L}$ in $S^{3}.$%
The crossing change operation depicted in Fig.9 which is equivalent to
elementary Dehn surgery was used by Hempel$^{93}$ in his proof of the
following\bigskip\ 

\textbf{Theorem 4.1}.(Hempel) \textit{If} M \textit{is a compact connected
orientable} 3-\textit{manifold}, \textit{then there is a collection} \{T$%
_{1} $,...,T$_{k}$\} \textit{of mutually exclusive solid tori in} M and a
collection \{\'{T}$_{1}$,...,\'{T}$_{k}$\}

\textit{of mutually exclusive solid tori in} $S^{3}$ \textit{such that the
closure of} (M$\smallsetminus $\{T$_{1}$,...,T$_{k}$\}) \textit{is
homeomorphic to the closure of} ($S^{3}\smallsetminus $\{\'{T}$_{1}$,...,\'{T%
}$_{k}$\}). \textit{Furthermore, the tori} \{\'{T}$_{1}$,...,\'{T}$_{k}$\}%
\textit{\ may be}

\textit{chosen in such a way that they are unknotted in }$S^{3}\bigskip $

\textbf{Remark 4.2.} Chronologically this theorem is precursor of Kirby
calculus. Generic examples of topologically different links related to each
other homeomorphically can be found on page 49 of$^{54}$ or in Fig. 12 of$%
^{22}.$Evidently, that the unknotedness of the collection of tori \ \{\'{T}$%
_{1}$,...,\'{T}$_{k}$\} is not essential. \ 

\textbf{Remark 4.3}. Since every 3-manifold is obtainable by surgery on some
link $\mathcal{L}$ it is possible (using Theorem 4.1.) to go backward in
this process and to reobtain $S^{3}$ with some link $\mathcal{L}^{^{\prime
}} $ related to $\mathcal{L}$ via some homeomorphism. Because the elementary
surgery can be modelled by the elementary crossing change (Fig.s 8,9) this
means that the link $\mathcal{L}^{\prime }$ should be simpler than $\mathcal{%
L}$. Accordingly, the Seifert surface for such a link should be simpler. It
should look either as a) Seifert surface for a collection of unknots, b)
Seifert surface for collection of trefoils or, c) Sifert surface for
collection of unknots and trefoils. E.g. see Fig.8.22 of Ref.66. More on
this will be said below, in the next subsection.

Supplied information brings us much closer to proving/disproving that $%
\Delta _{\text{K,M}}(t)=$ $\Delta _{\text{K, S}^{3}}(t).$ To answer this
question \ we need to recall and to use Theorems 3.13.(b) and 3.13 (c)\ and
13.15. \textsl{These theorems contain the essence of black magic of
concordance in} \textsl{the most concentrated form}. They are providing
information about how the homology sphere 3- manifold M should look like. It
is made of nontrivial knots and \ slice knots. Recall that the slice knots \
are representing the vacuum state while the nontrivial knots are
representing particles. The nontrivial knots eventually relax to stable
particles represented by trefoils. Thus, trefoils are living in $S^{3}.$ The
Dehn surgeries can be made on trefoils providing us with the hadron
excitation spectrum. Mathematically, such Dehn surgeries convert $S^{3}$
into the homology 3-sphere. \textsl{Remarkably, these are exactly the main
features \ defining the Casson invariant} defined in subsection 3.2!
According to Theorem 3.15 we notice that for homology spheres slice knots
are not linked with nontrivial knots. In mathematics such a situation is
characterized in terms of either \textsl{unlinks} or \textsl{boundary links.}

\textbf{Definition 4.4. }\textit{Let} $k$ \textit{and} $\QTR{sl}{l}$ \textit{%
be some knots \ forming a link} $k\cup $ $\QTR{sl}{l,}$ \textit{then the
unlink is characterized by the condition }$lk(k,l)=0.$

\textbf{Definition 4.5.} \textit{A link} $L$ $=L_{1}\cup \cdot \cdot \cdot
\cup $ $L_{n}$ \textit{is a} \textsl{boundary link} \textit{if there exists
an orientable Seifert surface }$S$ \textit{made of} $n$ \textit{disjoint
components} $S=$ $S_{1}$ $\cup \cdot \cdot \cdot \cup $ $S_{n}$ \textit{such
that for each} $i<n$ \textit{we} \textit{have} $\partial S_{i}=L_{i}$

Now we are in the position to formulate yet another\bigskip

\textbf{Theorem 4.6.(}Saveliev$^{56}$\textbf{,} page 94\textbf{)} \textit{Let%
} $k\cup $ $\QTR{sl}{l}$ \textit{be a boundary link in a homology sphere} $%
\Sigma ,$ \textit{and let} $\Sigma ^{\prime }=\Sigma +\varepsilon k,$ 
\textit{a surgery of} $\Sigma $ \textit{along} $k$ \textit{with} $%
\varepsilon =\pm 1.$ \textit{Then} $\Delta _{l\subset \Sigma }(t)=\Delta
_{l\subset \Sigma ^{^{\prime }}}(t)$ \textit{where} $l\subset \Sigma
^{\prime }$ \textit{is the image of} $l\subset \Sigma $ \textit{under the
surgery}.\bigskip

\textbf{Remark 4.7.} The content of this theorem is just a restatement of
Hempel's theorem in a specific setting. Saveliev's theorem is illustrated in
Fig.s 8.9 superimposed with Fig.12 of Ref.22. Indeed, in Fig.s 8,9 it is
sufficient to identify $k$ with the thickened unknot and $l$ with, say, a
trefoil.

Using Theorem 4.6. it is possible to prove the following \bigskip

\textbf{Theorem 4.8.(}Saveliev$^{56}$\textbf{, }page 95\textbf{)} \textit{Let%
} $k$ $be$\textit{\ a knot in a homology sphere} $\Sigma .$ \textit{Then} 
\textit{there exists a knot} $l$ \textit{in} $S^{3}$ \textit{such that} $%
\Delta _{k\subset \Sigma }(t)=\Delta _{l\subset S^{3}}(t).\bigskip $

\textbf{Remark 4.9. }Just stated theorem provides negative answer to the
question: "Could it be that $\Delta _{\text{K,M}}(t)=$ $\Delta _{\text{K, S}%
^{3}}(t)$ ?" This fact apparently creates some difficulty in calculating of $%
\Delta _{\text{K,M}}(t).$ The situation can be dramatically improved based
on the following observations.

Let us try to apply the definition of derivative in \ \ (3.20) to the
boundary link made out of two components K and L. Following$^{56}$ this can
be achieved by replacing (3.20) by the second derivative defined as follows:%
\begin{eqnarray}
&&\lambda (\text{K}_{m+1},\text{L}_{n+1};\text{M)-}\lambda (\text{K}_{m},%
\text{L}_{n+1};\text{M)-}\lambda (\text{K}_{m+1},\text{L}_{n};\text{M)+}%
\lambda (\text{K}_{m},\text{L}_{n};\text{M)}  \notag \\
&=&\lambda ^{\prime }(\text{K},\text{L}_{n+1};\text{M)-}\lambda ^{\prime }(%
\text{K},\text{L}_{n};\text{M)}  \notag \\
&=&\lambda ^{\prime }(\text{K}_{m+1},\text{L};\text{M)-}\lambda ^{\prime }(%
\text{K}_{m},\text{L};\text{M)}\equiv \lambda ^{^{\prime \prime }}(\text{K},%
\text{L};\text{M).}  \TCItag{4.4}
\end{eqnarray}%
If, following Casson, we require that $\lambda ^{^{\prime \prime }}($K$,$L$;$%
M)=0, this would imply that $\lambda ^{\prime }($K$_{m+1},$L$;$M)=$\lambda
^{\prime }($K$_{m},$L$;$M) and $\lambda ^{\prime }($K$,$L$_{n+1};$M)=$%
\lambda ^{\prime }($K$,$L$_{n};$M) which can be formulated as

\textbf{Corollary 4.10. }Although the Theorem 3.13 (b) does not mention
about the mutual alignment of slices and nontrivial knots, just obtained
result requires slices and nontrivial knots to be arranged as boundary
links. \ Only under such circumstances equation (3.31) is valid.

Because of equivalence between elementary Dehn surgeries and crossing
changes (Fig.s 8,9), it is essential now to investigate how the key equation
(3.30) should be modified to account for this equivalence. \ We begin with
(3.30) in which M$=S^{3}.$ Next, we look at the derivative (3.20) which, in
view of Fig.s 8,9, can be presented as follows%
\begin{equation}
\lambda ^{\prime }(\text{K},\text{S}^{3})=\lambda (\text{K};S^{3}+\mathcal{O}%
)-\lambda (\text{K};\text{S}^{3}).  \tag{4.5}
\end{equation}%
where $\mathcal{O}$ is the unknot concordant to slice knot. Next, the
property of n-independence of the derivative can be restated in view of
equation (4.4) as follows%
\begin{equation}
\lambda ^{\prime }(\text{K},\text{S}^{3})=\lambda (\text{K};S^{3}+\mathcal{O}%
_{1}\mathcal{+O}_{2})-\lambda (\text{K};\text{S}^{3}+\mathcal{O}%
_{1})=\lambda (\text{K};S^{3}+\mathcal{O}_{1})-\lambda (\text{K};\text{S}%
^{3})  \tag{4.6}
\end{equation}%
Evidently, this result implies 
\begin{equation}
\lambda (\text{K};S^{3}+\mathcal{O}_{1}\mathcal{+\cdot \cdot \cdot +O}%
_{n})=\lambda (\text{K};\text{S}^{3})+\frac{n}{2}\Delta _{\text{K},\text{S}%
^{3}}^{\prime \prime }(1).  \tag{4.8}
\end{equation}%
By design, equation (4.8) is equivalent to the equation (3.31). Just
obtained results require additional streamlining. This is subject matter of
the next subsection.\bigskip\ 

4.3. Fine structure of the physical vacuum. Magical role of
concordance\bigskip \bigskip

Up to now our readers were left under impression that only slice knots/links
represent the physical vacuum. However, it is known$^{44,74}$ that every
ribbon knot/link is slice knot/link but the converse remains only as
conjecture. In fact, there are many examples$^{99}$ of slice knots/links
which are are ribbons. An example of \ ribbon and slice knots is depicted in
Fig.11 \ 

\begin{figure}[ptb]
\begin{center}
\includegraphics[scale=1.4]{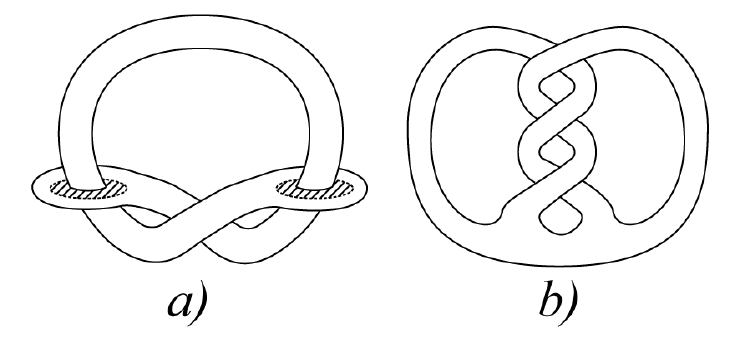}
\end{center}
\caption{ Example of the ribbon a) and slice b) knots}
\end{figure}

\ To make a ribbon is easy. Indeed, for any knot K and its mirror image $m$K
the connected sum K\# $m$K is a ribbon$^{44}$. Recall from section 3 that
Arf (K) is the concordance invariant. This means$^{72,100}$ that if K is
slice, then Arf (K)=0 . In view of equations (3.16), (3.31b) and (3.32) and
also taking into account the content of Theorem 3.13 (b) the result Arf
(K)=0 when K is slice knot/link makes perfect physical sense. The C-F
formula (3.32) describes the hadron mass spectrum of mesons/baryons starting
from the meson/baryon of lowest but nonzero mass. Since slice/ribbon
knots/links had been identified with the physical vacuum, we do not need to
worry about the mass spectrum coming from such a vacuum. According to
Kauffman$^{72}$ Arf (K) can acquire only two \ values: 0 for the vacuum,
represented, say, by the slice/ribbon knots, and 1, for all other knots.
These other knots can be reduced to the trefoil knot via band pass
operation. At the level of Seifert surface, e.g. see Fig.10, these pass band
operations preserving the pass class of the boundary of this surface
(because it is oriented) are depicted in Fig.12\ 

\begin{figure}[ptb]
\begin{center}
\includegraphics[scale=1.4]{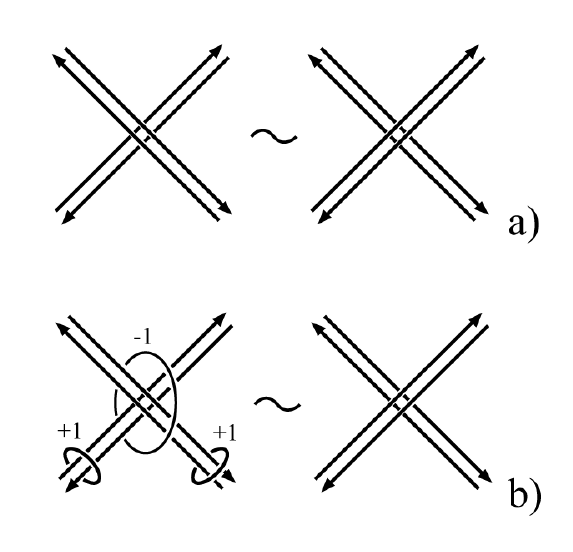}
\end{center}
\caption{ The band pass-move a) formally defined. The same move, but done
with help of Kirby moves b)}
\end{figure}

Thus, the Arf (K) is organizing the set of all knots/links K into two
classes. It generates the equivalence relation on $\mathcal{K}$-the totality
of all knots. In physics such a subdivision reflects the fact that massless
particles can never acquire nonzero rest mass. In the reminder of this
subsection we shall discuss processes leading to such a subdivision.

We begin with the following question. Is the operation of crossing changes
the only mechanism by which knots/links can be unknotted/unlinked? Said
alternatively ( if we \ use the surgery using Dehn twists): Is this the only
operation connecting a given state (that is 3-manifold, perhaps, associated
\ with framed nontrivial knot) to the vacuum state (that is 3-manifold,
perhaps, associated with nontrivial knot without framing)?

Following seminal papers by Levine$^{101}$ and Milnor$^{102}$ it is
convenient to introduce the concept of \textsl{link homotopy}.

\textbf{Definition 4.11}. \textit{\ Link homotopy} \textit{\ is an
equivalence relation generated by the ambient isotopy} \textit{and the
crossing change move}, \textit{when two branches of the crossing belong to} 
\textsl{the same link component}.

Immediately, the question emerges: \textsl{What equivalence is stronger:
surgical or homotopy}? To measure the degree of homotopy, Milnor suggested
his (now known as Milnor) invariants of link homotopy. These can be defined
as follows. Let the set of subscripts $\{i_{1},...,i_{k}\}$ be denoted as $%
\mathfrak{i.}$ Here all indices are distinct and the label $k$ refers to the 
$k$-th component of the link. Then the commonly accepted notation for the
Milnor $\bar{\mu}$-invariant is $\bar{\mu}_{\mathfrak{i}}.$ In particular,
for the two-component links $\bar{\mu}_{12}$ is a complete link homotopy
invariant. It is just the familiar linking number. For the three-component
links the collection made of $\bar{\mu}_{12},\bar{\mu}_{23},\bar{\mu}_{13}$
and $\bar{\mu}_{123}$ mod \{$\bar{\mu}_{12},\bar{\mu}_{23},\bar{\mu}_{13}\}$
is a complete set of link homotopy invariants. Evidently, Milnor's
-invariants generalize the concept of the linking number. Their power can be
seen when one is looking, for example, at the Borromeo link Fig.13 b) or
Whitehead link, Fig. 13a).

\begin{figure}[ptb]
\begin{center}
\includegraphics[scale=1.4]{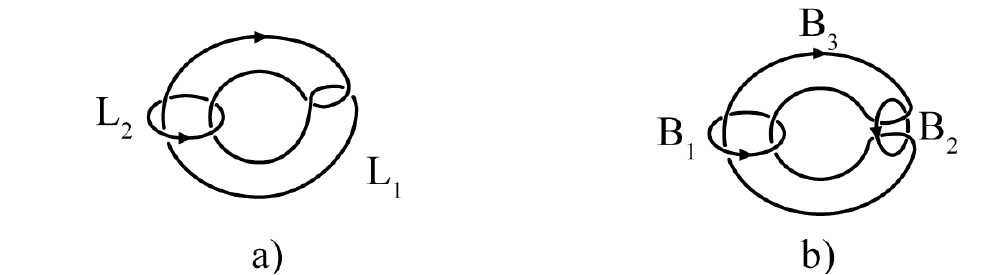}
\end{center}
\caption{ The Whitehead link a) and the Borromeo link b)}
\end{figure}

In both cases the linking numbers are zero while Milnor's numbers are not.
Levine demonstrated that only for two and tree component links the homotopy
equivalence is identical to the surgical equivalence. This is very deep and
nontrivial result. It generalizes Hempel's result depicted in Fig. 9.
Subsequently, it was demonstrated$^{103,104}$ that the link concordance
implies link homotopy. From here we obtain the triangle of equivalence
relations for two and three-component links in which the concordance is
playing the dominant role. Levine's results were reobtained independently by
Matveev$^{105}$ who studied surgical versus homotopy equivalence for
knots/links in the homology spheres. Matveev's results are of physical
importance as we would like to explain now. To do so, we notice that some
(but not all!) results of$^{105}$ are further refined in$^{106}$. Christine
Lescop published paper$^{107}$ containing considerable amount of details
missing in both$^{105,106}$ . Thus, using both Matveev's and Lescop's
results we inject some physics into them. The main result of Lescop can be
formulated as\bigskip

\textbf{Theorem 4.12.} \textit{Any integral homology sphere with Casson
invariant zero can be obtained from} $S^{3}$ \textit{by the sequence of} $%
\pm $1 Dehn surgeries on the boundary link $L$ each component of \textit{%
which} $L_{i}$ , $i=1,...,n,$ \textit{has a trivial Alexander polynomial.
That is} $\Delta _{L_{i}}(t)=1.\bigskip $

\textbf{Remark 4.13. }In$^{107}$ Lescop considers $\frac{1}{2}\Delta _{\text{%
K},\text{M}}^{\prime \prime }(1)$ as the Casson invariant of K.

This result by Lescop can be restated in physical language. For this we need
to introduce such technical concepts as "smooth concordance" group$\footnote{%
The most interesting physically}$ \ $\mathcal{H}_{diff}$ , "topological
concordance" group $\mathcal{H}_{top}$ and "algebraic concordance" group $%
\mathcal{H}_{a\lg }$. The $\mathcal{H}_{a\lg }$ is defined in$^{40,42}$ , $%
\mathcal{H}_{diff}$ and $\mathcal{H}_{top}$ are defined in$^{108}$ . Without
going into details , it can be demonstrated that $\mathcal{H}%
_{diff}\rightarrow \mathcal{H}_{top}\rightarrow \mathcal{H}_{a\lg }$ . At
the level of $\mathcal{H}_{top}$ Michael Freedman proved$^{109}$, Chr.11,
paragraph 7B, the following\bigskip

\textbf{Theorem 4.14.}(Friedman) \textit{Any knot with trivial Alexander
polynomial is slice in topological category}.\bigskip

\textbf{Remark 4.15.} All these distinctions originate from a delicate
difference in embeddings of a cone \^{k} (a disc $D^{2}$), Fig.1., into
4-dimensional ball B$^{4}$. These are issues associated with the topology of
4-manifolds. If we suppress these distinctions, leaving them to
mathematicians, then Theorems 4.12. and 4.14. can be reformulated in the
physical language as follows:\bigskip

\textsl{Scattering processes involving stable massive particles are not
affected by the vacuum fluctuations.\bigskip\ }Alternatively: \textsl{%
Massless particles cannot be described in terms of the C-F plot\bigskip
\bigskip }

Thus, the sequence of different homology spheres can be created \ by either
the Dehn surgery on boundary links or the Dehn surgery on a knot K in $S^{3}$%
. Matveev$^{105}$ demonstrated that $\pm $1 Dehn surgeries on Borromeo rings
in $S^{3}$can be also used to create the homology spheres. It is of interest
to provide some specifics (without proofs). Instead of Dehn surgeries
Matveev considers a combination of crossing changes and Kirby moves (e.g.
see Fig.s 8,9) on the boundary links. He describes a bit different type of
surgery (as compared with the protocol described in Appendix ). It is
depicted in Fig.14.\ 

\begin{figure}[ptb]
\begin{center}
\includegraphics[scale=1.5]{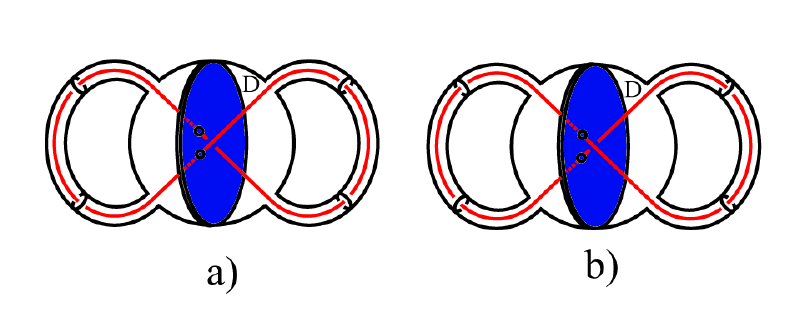}
\end{center}
\caption{ The 1st Matveev move (he is calling it the "Whitehead move")}
\end{figure}

The initial state of the handlebody of genus 2 (with the individual sublink)
differs from the final state by cutting it along the separating curve (a
disc), twisting it by 360$^{0}$; and regluing it back. Instead of the Dehn
surgery protocol, Matveev's surgery protocol is fully complacent with
Milnor's definition of homotopy transformations and is described as follows.
1.Take a handlebody of genus 2 ( along with the sublink/ link lying inside)
out of $S^{3}$. 2.Make just described 360$^{0}$ twist (not to be confused
with the Dehn twist!). 3.Reglue. 4.Insert such modified handlebody back into
S$^{3}$. The net result produces the effect which is equivalent to that
depicted in Fig.s 8,9. Therefore, Matveev's surgery is equivalent to the
more traditional Dehn surgery! What we are gaining if we use Matveev
prescription for the surgery instead of the traditional one? It happens that
such described modification of surgery is very helpful since it allows an
extension to the surgery on Borromeo rings. Matveev distinguishes between
the strict Borromeo surgery, when it is performed distinctly on three rings
and non strict surgery, when it is performed on lesser than three rings.
Each non strict Borromeo surgery can be made out of superposition of strict
Borromeo surgery and Kirby moves, Matveev claims. Furthermore, he claims
that transformations depicted in Fig.14 can be achieved with help of
Borromeo surgeries too! Skipping details which can be found in$^{105}$ , the
net result of strict Borromeo surgery is depicted in Fig.15.\ 

\begin{figure}[ptb]
\begin{center}
\includegraphics[scale=1.2]{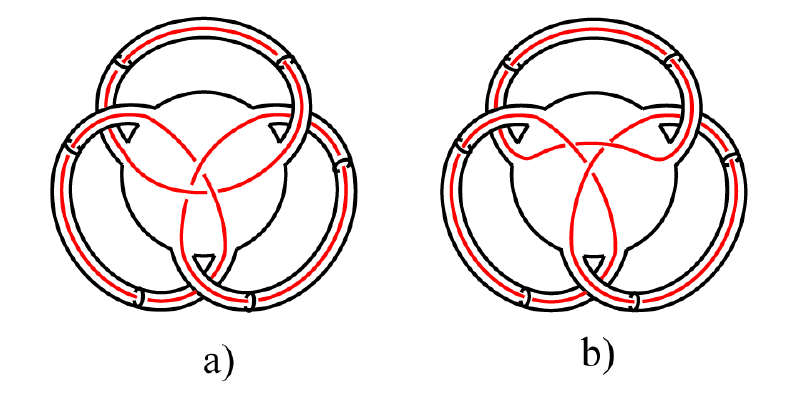}
\end{center}
\caption{The 2nd Matveev move (he is calling it the"strict Borromeo
surgery") }
\end{figure}

Since the crossing change is equivalent to the Dehn surgery, as Matveev
argues, the Borromeo surgeries, depicted in Fig.15, can be used for creation
of homology spheres. However, from$^{54}$ we know that that the Borromeo
rings are \textbf{not} boundary links! Thus, it seems like we cannot apply
Theorems 4.12 and 4.14. to Borromeo rings. Furthermore, we cannot apply
Theorem 3.13(b) as well. Very recently Krushkal$^{110}$ demonstrated
(following ideas of Michael Friedman) that, nevertheless, in a certain
(4-dimensional) sense the Borromeo rings are slice and, therefore, Theorem
3.13(b) can be actually applied! Because of this recent discovery, we arrive
at the same physical conclusions as stated immediately after Remark
4.15.\bigskip \bigskip

\textbf{Remark 4.16.} While \ the description of Fig.14 provides explicitly
details of crossing change operation, Fig.15 provides only the initial and
the final stages of the so called $\Delta -$move operation\footnote{%
This terminology is taken from the theory of finite type invariants. In knot
theory such a move is easily recognizible as the 3rd Reidemeister move.}
Details are missing in Matveev's paper. These details were provided for the
first time in the paper by Goussarov$^{111}$. In literature one can
encounter statements that in Habiro's paper$^{112}$ the same results were
obtained. This is not true, however, since the paper by Habiro does not
contain the description of $\Delta -$move at the level of handlebody of
genus 3. Habiro's paper is more combinatorial and algebraic while
Goussarov's is more topological.\bigskip

Combination of Kirby and $\Delta -$moves are discussed in detail in$%
^{98,113} $. All these papers lie at the foundation of theory of finite type
invariants$^{50,75}$ . Since this theory goes far beyond the homology sphere
case, there is no need to discuss it here. This theory uses essentially both
the Kontsevich integral and the Chern-Simons functional. These are being
used perturbatively. As result, the output is given in the form of Theorem
3.16 of section 3. This is consistent with the definition given by Lescop
(Remark 4.13) where $\frac{1}{2}\Delta _{\text{K},\text{M}}^{\prime \prime
}(1)$ was used instead of $\frac{1}{2}\Delta _{\text{K},\text{S}%
^{3}}^{\prime \prime }(1)$ since theory of finite type invariants is not
restricted to the homology sphere case. Physically, however, we believe \
that the homology sphere case discussed in detail in this paper is
sufficient. In addition, the existing theories of finite type invariants are
all perturbative to our knowledge while, in our opinion, the Floer-style
nonperturbative approach$^{16,17}$ is essential for detecting fine details
not detectable in perturbative treatments.\bigskip

4.4. Concordance at the next level. Doubly sliced knots and links\bigskip

From previous discussion it follows that the notion of concordance and of
slice (knot) are synonymous. The equivalence relation of concordance
originated in early 1960's in works of Fox, Kerwaire and Milnor in the
context of theory of isolated singularities of 2-spheres in 4-manifolds. Let 
$\mathcal{K}$ be the set of ambient isotopy classes of knots and let $%
\mathcal{C}$ the set of(smooth) concordance classes of knots. Since isotopic
knots are concordant$^{103,104}$, there is a natural surjection $\mathcal{K}$
$\rightarrow $ $\mathcal{C}$. As we already demonstrated , the connected sum
operation \ endows $\mathcal{C}$ with the structure of an Abelian group
called \textsl{smooth knot concordance group}. The identity element is
unknot. Any element concordant to unknot is called slice knot as we already
know. We also know that the slice knot is that knot which bounds a disc $%
D^{2}$ smoothly embedded in 4-ball B$^{4}$ (e.g. see Fig.1). Surprisingly, a
systematic extension$^{114}$ of the notion of concordance to links was made
only in 2012! Since the notion of concordance is related to topology of
4-manifolds, it should not come as total surprise that it is possible to
elevate this notion to the next level by introducing the notion of \textsl{%
doubly slice }knots (and, accordingly, of \textsl{double concordance)}.
Whether or not doubly sliced knots/links have any relevance to physical
reality remains to be investigated. Nevertheless, it is important to be
aware of their existence already at this level of presentation. Thus, knot
is called slice if it can be realized as the equator of an embedding of $%
S^{2}$ into $S^{4}$. Accordingly, a knot is \textsl{doubly slice} if it can
be realized as the equator of an unknotted embedding of $S^{2}$ into $S^{4}$
(recall that the "classical" knot K is an embedding of $S^{1}$ into $S^{3}$%
). Understanding when knots are slice or doubly slice, is helpful for our
understanding of 4-dimensional topology in more familiar setting of knots in 
$S^{3}$. One of the most fascinating aspects of topology of 4-manifolds is
the discrepancy between the smooth and topologically locally flat spaces
(e.g. read previous subsection). This discrepancy is intrinsic only for
4-manifolds$^{43,115}$ . It is nicely captured in study of slice knots$%
^{116} $ . Recently, Meier proved the following\bigskip

\textbf{Theorem 4.17}. (Meier$^{117}$) \textit{There exists an infinite
family of slice knots that are topologically doubly slice, but not smoothly
doubly slice}.\bigskip \bigskip

This paper is part of Meier PhD thesis$^{118}$. His proofs involve results
from the Heegard Floer homology. A quick introduction to this field of study
is given ch-r 9 of . In connection with Theorem 4.17. several questions
arise: Is infinite family of slice knots in this theorem made of slice knots
identifiable with the physical vacuum? If the answer is "yes", what this non
smoothness means physically? If the answer is "no", what else these slice
knots represent? Providing answers to these questions based on physical
considerations might help to resolve the deepest mysteries of 4-manifolds
topology, e.g. those discussed in$^{43,109}$.\bigskip \bigskip

{\Large 5. \ }{\large Brief summary and discussion\bigskip }

Some time ago Ranada obtained new nontrivial solutions of the Maxwellian
(that is the Abelian gauge) fields without sources which were reinterpreted
in our work $^{15}$ as particle-like (monopoles, dyons, etc.). These
solutions were obtained by the method of Abelian reduction of the
non-Abelian Yang-Mills functional. The developed method involved\ uses of
instanon-type calculations typically used for\ the non-Abelian gauge fields$%
^{13}$. \ Employing the electric-magnetic duality permits then to replace
all charges \ by the corresponding particle-like solutions of the
source-free \ Abelian gauge fields. Such a replacement has many advantages.
In particular, if one believes that the non-Abelian gauge fields are just
generalizations of more familiar Maxwellian fields, then one encounters a
problem about existence of non-Abelian charges -analogs of (seemingly
familiar) macroscopic charges in Maxwell's electrodynamics. Surprisingly, to
introduce the macroscopic charges into non-Abelian fields is a challenging
task which (to our knowledge) is still not completed. By treating gravity as
gauge theory, the analogous problem emerges in gravity too. It is the
problem of description of motion of the extended bodies in general
relativity. \ The difficulties in description of extended objects in both
the Y-M and gravity \ fields are summarized on page 97 of Ref.$18$. The
method of Abelian reduction\ developed in$^{15}$ and further extended in$%
^{22}$ allows us to improve the existing thus far situation dramatically by
employing for both the Abelian and non-Abelian gauge fields the same
methods. These were originally used only for description of non-Abelian
monopoles, dyons, etc., by employing source-free \ non-Abelian gauge fields.
Clearly, under such conditions, the problem of relating the source-free
non-Abelian configurations of gauge fields to their Abelian counterparts
with extended sources was left unexplained.

The newly developed Abelian reduction of the source-free gauge fields
resulted in dynamically generating (in the bulk) of all kinds of knot and
link structures excluding those made of hyperbolic knots/links.\ Next
logical steps in this development \ should involve study of likely
connections of the obtained knotted/linked structures with the hadron
physics and, more broadly, with the Standard Model and gravity. This paper
is the first logical step in this direction. To make our presentation self
-contained, we would like now to mention some recent (and not so recent)
works providing independent support to the results \ of this paper. The
story begins with influential paper by Atiyah and Manton entitled "Skyrmions
from instantons"$^{119}$. The Skyrme model is well studied model \ for
description of baryons$^{120}.$ Whether or not it can be applied to hadrons
was studied in$^{121}$ with positive outcome. \ The latest PhD work by
Jennings$^{122}$ convincingly demonstrates how skyrmions can be represented
by knots and links. These knots and links happen to be exactly of the type
described in detail in our work$^{22}.$ Ref.$123$ contains up to date review
of \ efforts by other researches.

Connections with sting models can be developed in principle by studying Ref.$%
124$ as point of departure. In it the problem of accounting for the presence
of topological constraints in systems such as entangled polymers, etc.,
undergoing some dynamics,$^{125}$ is studied. The work uses results of
Moffatt,$^{92}$ discussed in this work in section 4.1.,superimposed with
some results from the theory of reaction -diffusion (birth-death type)
processes. The authors of$^{124}$ do notice a connection between dynamics of
reaction-diffusion processes and dynamics of spin chains. In Ref.s $126,127$
not only \ the connections between reaction-diffusion processes and spin
chains were investigated, but in addition they were used as an input for
building several (topological) string models.

We mentioned already works by Taubes$^{16}$ and Masataka$^{17}$ in which
Floer-type approach to instantons (e.g. read Ref.15 for an introduction to
this approach) was used. In addition, Lim$^{128}$ demonstrated equivalence
between Seiberg-Witten and Casson invariants for homology 3-sphere.
Furthermore, Kronheimer and Mrowka using instanton Floer homology$^{129}$
reobtained Alexander polynomial used essentially in this work. \ Since
skyrmions are instantons and since knotted/linked skyrmions describe
hadrons, \ the instanton origin of Alexander polynomial provides needed
missing link between baryons and instantons. Details of just noticed
correspondences will be investigated in future publications.

\bigskip

\bigskip

\textbf{Appendix. Basics facts about the Dehn surgery and Kirby calculus}

\bigskip

1. \textsl{Dehn surgery}. Let K be some knot in 3-manifold M. We surround K
by a regular tubular neighborhood $N$(K). Evidently, if the boundary $%
\partial N$(K$)=T$ ,where $T$ is the torus $T$,

then $N$(K) \ the solid torus $N$(K)$=S^{2}\times D^{2}$ while $\mathring{V}$%
(K) is the solid torus without boundary. Following Rolfsen$^{54}$, suppose,
we are given: \ 

a) a 3-manifold M, perhaps with boundary. Let its interior be \r{M};

b) a link L= K$_{1}\cup \cdot \cdot \cdot \cup $ K$_{n};$

c) a disjoint tubular neighborhoods $N($K$_{i})\equiv N_{i}$ of the K$_{i};$

d) a specified simple (that is without self intersection s) closed curve $%
J_{i}$ on each $\partial N$(K$_{i}).$

Given these data, we may construct the 3-manifold M$^{\prime }$%
\begin{equation*}
\text{M}^{\prime }=(M-(\mathring{N}_{1}\cup \cdot \cdot \cdot \cup \mathring{%
N}_{n}))\cup _{h}((N_{1}\cup \cdot \cdot \cdot \cup N_{n})),
\end{equation*}

where $h$ is a union of homeomorphisms $h_{i}:\partial N_{i}\rightarrow
\partial N_{i}\subset $ M, each of which take a meridian curve $\mu _{i}$ of 
$\partial N_{i}$ into the specified curve $J_{i}$.

\textbf{Definition A.1}. \ M$^{\prime }$ is said to be \ the result of a
Dehn surgery on M along the link L with \textsl{surgery instructions} c) and
d).\bigskip

In the most studied case M$=$S$^{3}$ or R$^{3}$ the surgery instructions are
being expressed by assigning a rational number $r_{i\text{ }}$ (could be $%
\infty $ sometimes) to each component of L$_{i}$

\textbf{Definition A.2.} Let $N($K$)=S^{2}\times D^{2},$ then a specified
homeomorphism $h_{i}:N($K$_{i})\rightarrow N($K$_{i})$ is called \textsl{%
framing} of $N($K$_{i}).$

If $J_{i}\subset \partial N$(K$_{i})$ is a simple closed curve, then the
following are equivalent

a) $J_{i}$ is a longitude $\lambda _{i}$ of $N($K$_{i});$

b) $J_{i}$ represents a generator of H$_{1}$($N_{i}$)$\cong \pi
_{1}(N_{i})\cong \mathbf{Z;}$

c) $J_{i}$ intersects some meridian of $N_{i}$ transversely in a single
point.

The meridian is intrinsic part of $N_{i}$ while the longitude $\lambda _{i}$
involves a choice.

\textbf{Definition A.3. }A \textsl{preferred framing} for $N_{i}$ is such
for which $\lambda _{i}$ is oriented in the same way as K$_{i}$ and the
meridian $\mu _{i}$ has linking number $\pm 1$ with K$_{i}.$ In such a case 
\begin{equation*}
h(\mu _{i})=[J_{i}]=a_{i}\lambda _{i}+b_{i}\mu _{i}.
\end{equation*}

Here $a_{i}$ is determined with accuracy up to a sign (dependent upon
orientation of $J_{i}$) while $b_{i}=lk($K$_{i}$,$J_{i}).$

\textbf{Definition A.4.} The ratio 
\begin{equation*}
r_{i\text{ }}=b_{i}/a_{i}
\end{equation*}

is called\textsl{\ surgery coefficient associated with} K$_{i}.$ If $a_{i}=0$
then, by definition, $b_{i}=\pm 1$ and $r_{i\text{ }}=\infty .$ Clearly, in
this case the surgery is trivial.

\textbf{Definition A.5. }The surgery is called \textsl{integral }if $%
a_{i}=\pm 1.$

Any link L with rational numbers attached to its components determines a
surgery yielding closed oriented 3-manifold. All closed oriented 3-manifolds
arise in this way. In knot theory it is common practice to draw 3-manifolds
by drawing the corresponding link with surgery coefficients $r_{i\text{ }}$%
placed next to respective components. \ If $U$ is unknot, then $r=0$ defines
a Lens space L(0,1), while for $r=\pm 1,\pm 1/2,\pm 1/3,...$ surgery on $U$
yields back $S^{3}$ . The first nontrivial surgery leading to the Poincar$%
e^{\prime }$ homology sphere is obtained by performing $\pm 1$ surgery on
the trefoil knot. The same result can be achieved by performing surgery on
many other links which topologically are \textbf{not} the same$^{60}$. This
observation lies at the heart of Kirby calculus.

2. \textsl{Kirby calculus and physics associated with them. }We just
discussed the fact that the Poincar$e^{\prime }$ homology sphere is
obtainable via Dehn surgery on the trefoil knot in $S^{3}.$ Fact of the
matter is that all 3-manifolds can be obtained surgically. As long as this
is done with links, the end result can be achieved in many ways while the
theorem of Gordon and Luecke$^{32}$ requires 1-to-1 correspondence between
knots and their complements in $S^{3}$. Therefore, by performing surgery on
knots naively, we should expect again 1-to-1 correspondence. This is not the
case however. In the previous subsection we noticed that the same Poincar$%
e^{\prime }$ homology sphere is obtainable either via surgery on the trefoil
knot or via surgery on many links as discussed in$^{60}$. Now, if knots are
associated with masses, then Gordon and Luecke theorem guarantees 1-to-1
correspondence between masses and 3 manifolds. This theorem guarantees that
the mass spectrum is discrete.

At this point we need to recall some facts from Einsteinian theory of
gravity. The original formulation by Einstein makes heavy emphasis on study
of the Schwarzshield solution of Einstein equations. These are the field
equations without sources in which the mass enters as an adjustable
parameter at the end of calculations. \ If masses are associated with knots,
we \ run into the same problems as in Einsteinian relativity. Specifically,
suppose that we have several masses, then the space around given knot, say,
trefoil, \ is made not only of the complement of the trefoil but out of
complements of the rest of knots/links existing in the trefoil complement.
Details are given in$^{22}.$Furthermore, following Einsteinian logic, we
shall assume that knots/links other than trefoil are closed geodesics in the
3-manifold created by the trefoil knot. In such a case, very much like in
Einsteinian gravity, we should assume that one can ignore the \ finiteness
of masses moving on geodesics. To account for finite masses and for extended
sizes of particles is always a great challenge$^{18}$, page 97.\ In section
2 we discussed how such a challenge can be plausibly resolved. It is also
difficult problem in knot theory$^{22,130}$ , where it is also not solved
systematically. This is so because Gordon and Luecke theorem is valid only
for complements of knots in $S^{3}$. Presence of other knots makes Gordon
and Luecke theorem non applicable. From this observation the following
question arises.

\textbf{Question A.6.} Suppose in some 3-manifold M (other than $S^{3})$
there are two knots K$_{1}$ and K$_{2}$ so that the associated complements
are M$\smallsetminus $K$_{1}$ and M$\smallsetminus $K$_{2}.$Suppose that \
there is a homeomorphism $h$: \ $h($M$\smallsetminus $K$_{1})=$M$%
\smallsetminus $K$_{2}.$ Will such a homeomorphism imply that K$_{1}=$K$_{2}$
? (With links this happens all the time. E.g. read Remark 4.2.)

If the answer to the above question is negative, this then would imply that 
\textsl{different} knots would have \textsl{the same complement} in M. \ If
we are interested in attaching some physics \ to these statements then, we
should only look for situations analogous to $S^{3}.$ This means that 
\textsl{all physical processes should be subject to selection rules }making
such degenerate cases physically forbidden. These selection rules presuppose
that the degeneracy just described can be realized in nature. And indeed, it
can! This leads to the concept of \textsl{cosmetic knots}. We refer our
readers to our paper$^{22}$, section 7, for further details.

The purpose of Kirby calculus is to establish the equivalence classes of
different links yielding the same 3-manifolds surgically. Deep down Kirby
calculus should be done using theory of 4-manifolds$^{43,97}$. Fortunately,
for the purposes of this work this path is not needed. In fact, as it was
rigorously demonstrated by Rolfsen$^{131,132}$, the \ four dimensional
approach to Kirby calculus could be entirely avoided. While Kirby calculus
use only integral surgery, Rolfsen's calculus use both the integral and
rational surgeries indiscriminately. Kirby calculus, however, are bit more
physically suggestive. This is so because of the following. \ The linking
number between two links $L_{1}$ and $L_{2}$ \ can be definied via study of
links projection on the arbitrary plane. Using Fig.4 if we identify arrows
with different links then we can prescribe $\varepsilon =1$ to K
configuration and $\varepsilon =-1$ to \={K} configuration so that 
\begin{equation*}
lk(L_{1},L_{2})=\dsum\limits_{i}\varepsilon _{i}
\end{equation*}%
where summation is over all crossings on the planar link diagram. In the
case of integral surgery we have $[J_{i}]=\lambda _{i}+b_{i}\mu _{i}$ . This
means that \ $J$-curve is making exactly one revolution in the direction
parallel to $\lambda _{i}.$ This can be illustrated \ as follows. Make a
ribbon link out of K$_{i}$ and $[J_{i}]$ as depicted in Fig.16 a)\ 

\begin{figure}[ptb]
\begin{center}
\includegraphics[scale=1.3]{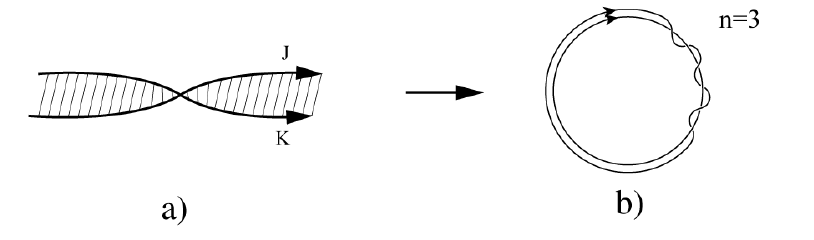}
\end{center}
\caption{ An exampmple of framing/self-linking}
\end{figure}

The self-linking number is determined then by $lk($K$_{i},\mu _{i})$, e.g.
see Fig.16b).\ Alternatively, we can introduce

\textbf{Definition A.6. \ }The\textsl{\ integral framing} of the knot K$_{i}$
corresponds to a choice $lk($K$_{i},\mu _{i})=b_{i}=n_{i}$ where $%
n_{i}=0,\pm 1,\pm 2,...$

An example of integral framing is depicted in Fig.16 b). Being armed with
these definitions we are ready to address the main problem studied by Kirby$%
^{120}$ : How to determine when two differently framed links produce the
same 3-manifold?

The answer to this question is given in terms of two (Kirby) moves. By
design, they are meant to establish the equivalence relation between links
with different framings producing the same \ 3-manifold. Using results of
our previous works$^{15,22}$, the first Kirby move, depicted in Fig.17 can
be interpreted physically as graphical statement illustrating the charge
conservation.\ 

\begin{figure}[ptb]
\begin{center}
\includegraphics[scale=1.3]{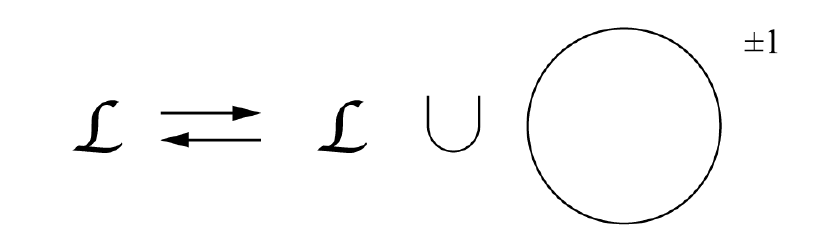}
\end{center}
\caption{ The 1st Kirby move}
\end{figure}

This move should be interpreted as follows: It is permissible to add or to
delete an unknot with framing $\pm 1$ which does not intersect other
components L$_{i}$ to a given link $\mathcal{L}$.

\begin{figure}[ptb]
\begin{center}
\includegraphics[scale=1.3]{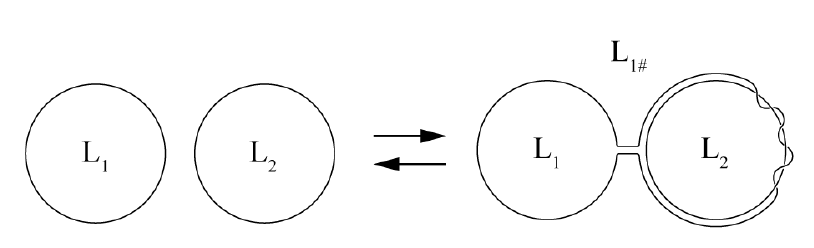}
\end{center}
\caption{ The 2nd Kirby move}
\end{figure}

The 2nd Kirby move is depicted in Fig.17 and can be physically reinterpreted
in terms of interaction between charges.\ Mathematically, this move can be
interpreted as follows. Let $L_{1}$ and $L_{2}$ be two link components
framed by integers $n_{1}$ and $n_{2}$ respectively and $L_{2}^{\prime }$ be
a longitude defining the framing of $L_{2}$ that is $lk(L_{2},L_{2}^{\prime
})=n_{2}.$ Replace now the pair $L_{1}$ $\cup $ $L_{2}$ by another pair $%
L_{1\#}$ $\cup $ $L_{2}$ in which $L_{1\#}=L_{1}\#_{b}L_{2}^{\prime }$ and $%
b $ is 2 sided band connecting $L_{1}$ with $L_{2}^{\prime }$ and disjoint
from another link components. While doing so, the rest of the link $\mathcal{%
L}$ remains unchanged. The framings of all components, except $L_{1},$ are
preserved while the framing of $L_{1}$ is changed into that for $L_{1\#}$
and is given by $n_{1}+n_{2}$ +$2lk(L_{1},L_{2}).$ The computation of $%
lk(L_{1},L_{2})$ proceeds in a standard way as described above (provided
that both $L_{1}$ and $L_{2}$ are oriented links).\bigskip \bigskip \bigskip

\textbf{Acknowledgement. }The author would like to thank the Editors of
IJMPA for careful selection of the referee for this paper. The author had
benefitted immensely from \ both the suggestions of the anonymous referee
and \ from countless conversations with Professor Louis Kauffman (University
of Illinois at Chicago) whose interest in subject matters, whose willingness
to help and whose insistence on precision in presenting the arguments had
brought this paper into its current form.

\bigskip

\bigskip {\large References}

1. \ \ \ M. Atiyah, N. Manton and B.Schroers, \textit{Proc.Roy.Soc}.\textit{A%
} \textbf{468, }1252 (2011).

2. \ \ \ R. Buniy, J. Cantarella, Th. Kephart and E. Rawdon,

\ \ \ \ \ \ \textit{Phys.Rev.D} \textbf{89,} 054523 (2014).

3. \ \ \ L. Faddeev and A.J. Niemi, \textit{Nature} \textbf{387}, 58 (1997).

4. \ \ \ D. Forster, \textit{J. High Energy Phys}. \textbf{2012}, 81 (2012).

5.\ \ \ \ K-I.\ \ Kondo, \ A. Ono, A. Shibata, T. Shinohara and T. Murakami,

\ \ \ \ \ \ \ \textit{J.Phys.A} \textbf{39}, 13767 (2006).

6. \ \ \ P. Baal, and A. Wipf, \textit{Phys.Lett.B} \textbf{515}, 181
(2001).\ \ \ \ \ \ 

7. \ \ \ E. Langmann and A. Niemi, \textit{\ Phys.Lett.B} \textbf{463}, 252
(1999).

8. \ \ \ Y. Cho, \textit{Phys.Lett.B} \textbf{644}, 2008 (2007).

9.\ \ \ \ K-I.\ Kondo, S. Kato, A. Shibata and T. Shinohara, arXiv:1409.1599.

10. \ W-S.\ Hou and G-G. Wong, \textit{\ Phys.Rev.D} \textbf{67,} 034003
(2003).

11. \ F. Brau, C. Semay, \textit{Phys.Rev.D} \textbf{72}, 078501 (2005).

12.\ \ V. Mathieu, N.Kochelev and V. Vento, \ \textit{Int.J.Mod.Phys.E} 18,1
(2009).

13. \ S. Donaldson, \textit{Floer Homology Groups in Yang-Mills Theory}

\ \ \ \ \ \ (Cambridge U. Press, Cambridge, UK, 2002).

14. \ P. Kronheimer, Th. Mrowka,\textit{\ Monopoles and Three-Manifolds}

\ \ \ \ \ \ \ (Cambridge U. Press , Cambridge UK, 2007).

15. \ A.\ Kholodenko, \textit{Analysis and Math.Phys}. (in press, available
online)

\ \ \ \ \ \ \ http://link.springer.com/article/10.1007\%2Fs13324-015-0112-6

\ \ \ \ \ \ arXiv:1402.1793.

16. \ C.Taubes, \textit{J.Diff.Geom}. \textbf{31}, 547 (1990). .

17. \ K.Masataka, \textit{Topology and Applications }\textbf{112, }111
(2001).

18. \ A.Kholodenko, \textit{Applications of Contact Geometry and Topology }

\ \ \ \ \ \ \textit{in Physics }(World Scientific Publishing Co.,Singapore,
2013).

19. \ H.K. Moffatt, \textit{\ J.Fluid Mech.} \textbf{159}, 359 (1985).

20. \ J. Etnyre, R. Ghrist, \textit{Transactions AMS} \textbf{352}, 5781
(2000).

21. \ A. Enciso, D. Peralta-Salas, \textit{Annals of Math}. \textbf{175},
345 (2012).

22. \ A.Kholodenko, \textit{Annals. of Phys}. (submitted),

\ \ \ \ \ \ arXiv: 1406.6108.

23. \ E.Witten, \textit{Comm.Math.Phys. }\textbf{121}. 351 (1989).

24. \ M.Atiyah, M., \textit{The Geometry and Physics of Knots},

\ \ \ \ \ \ (Cambridge University Press, Cambridge, UK, 1990).

25. \ Y.Cho, \textit{Phys.Lett.B} \textbf{644,} 208 (2007).

26. \ Y.Cho. F.Cho and J.Yoon,

\ \ \ \ \ \ \textit{Class.Quantum Grav}.\textbf{30} 055003 (2013).

27. \ M.Hutchins, \textit{Bull. AMS} \textbf{47},73 (2009).

28. \ N. Zung, A. Fomenko, \textit{Russian Math.Surveys} \textbf{45,}109
(1990).

29. \ J.Birman and R.Williams, \textit{Topology}\textbf{\ 22},47 (1983).

30. \ R.Ghrist, \textit{Topology} \textbf{36}, 423 (1997).

31. \ J.\ Birman, R. Williams, \textit{Cont. Math.} \textbf{209},1 (1983).

32. \ C. Gordon, J. Luecke,\textit{\ J.Amer.Math.Soc.}\textbf{\ 2}, 371
(1989).

33. \ W.Thurston, \textit{Bull. AMS} \textbf{6}, 367 (1982).

34. \ A.Kholodenko,\textit{\ J.Geom.Phys}.\textbf{\ 58}, 259 (2008).

35. \ S.Kamada, \textit{Braid and Knot Theory in Dimension Four}

\ \ \ \ \ \ \ (AMS Publishers, Providence, RI, 2002).

36. \ S.Carter, S.Kamada and M Saito, \textit{Surfaces in 4-Space}

\ \ \ \ \ \ \ (Springer-Verlag, Berlin, 2004).

37. \ Y. Choquet-Bruhat, \ General Relativity and Einstein Equations,

\ \ \ \ \ \ \ ( Clarendon Science Publications, Oxford, 2009).

38. \ \ M.Khuri, The positive mass theorem of general relativity,

\ \ \ \ \ \ \ www.math.sunysb.edu

39. \ R.Bartnik, Int. Congress in Mathematics \textbf{3}, 1 (2002).

40. \ C.Livingston, S.Naik, \textit{Introduction to knot concordance} (work
in progress)

\ \ \ \ \ \ \ wolfweb.unr.edu

41. \ K.Murasugi, \textit{Knot Theory and Its Applications}

\ \ \ \ \ \ \ (Birkh\"{a}user, Boston, MA, 1996).

42. \ J.Collins, \textit{On the concordance orders of knots},

\ \ \ \ \ \ \ PhD thesis (Department of Mathematics, U. of Edinburgh, UK,
2011).

43. \ A.Scorpan, \textit{The Wild World of 4-Manifolds}

\ \ \ \ \ \ \ (AMS Publishers, Providence, RI, 2005).

44. \ L.Kauffman, \textit{On Knots}

\ \ \ \ \ \ (Princeton U. Press, Princeton, 1987).

45. \ S.Akbulut and J.McCarthy, \textit{Casson Invariant for Oriented
Homology}

\ \ \ \ \ \ \ \textit{3-spheres. An Exposition} (Princeton U.Press,
Princeton, NJ 1990).

46. \ S.Elliott and M.Frang, \textit{Rev. Mod.Phys.} \textbf{87}, 137 (2015).

47. \ R. Feynman, R.Leighton and M.Sands, \textit{Feynman Lectures on
Physics,}

\ \ \ \ \ \ \ Vol.3 (Addison-Wesley Publ.Co, Reading, MA 1965).

48. \ V.Manturov, \textit{Knot Theory} (CRC Press LLC, Boca Raton, FL, 2004).

49. \ L. Kauffman,\textit{\ Eur.J.Combinatorics} \textbf{20}, 663 (1999).

50.\ \ S. Chmutov, S. Duzhin and J. Mostovoy, \textit{Introduction to
Vassiliev Knot Invariants}

\ \ \ \ \ \ (Cambridge U.Press, Cambridge, 2012).

51. \ Ch. Livingston, \textit{Knot Theory} (Mathematical association of
America,

\ \ \ \ \ \ \ Washington DC, 1993).

52.\ \ J.Weeks, \textit{Notices \ AMS} \textbf{51},610 (2004).

53. \ R.Aurich, S.Lustig, \textit{Class.Quantum Grav}. \textbf{31}, 165009
(2014).

54. \ D. Rolfsen, \textit{Knots and Links} (Publish or Perish, Houston TX,
1976).

55. \ N.Saveliev,\textit{\ Invariants for Homology 3-Spheres}

\ \ \ \ \ \ \ (Springer-Verlag, Berlin, 2002).

56. \ N.Saveliev, \textit{Lectures on the Topology of 3-manifols}, 2nd
edition,

\ \ \ \ \ \ \ (Walter de Gruyter, Berlin, 2012).

57. \ P.Kronheimer, T.Mrowka, \textit{Geom.\&Topol}.\textbf{8}, 296 (2004).

58. \ S.Boyer, \textit{Chaos, Solitons and Fractals} \textbf{9}, 657 (1998).

59. \ C.McA Gordon, Knots, homology spheres, and contractible 4-manifolds,

\ \ \ \ \ \ \textit{Topology} \textbf{14}, 151 (1975).

60. \ V.Prasolov, A. Sossinsky, \textit{Knots, Links and 3-Manifolds}

\ \ \ \ \ \ (AMS Publishers, Providence, RI, 1997).

61. \ R.Fintushel, R.Stern, Knots, links and 4-manifolds,

\ \ \ \ \ \ \textit{Invent. Math}. \textbf{134}, 363 (1998).

62. \ D.McDuff, D.Salamon, \textit{Introduction Into Symplectic Topology}

\ \ \ \ \ \ (Clarendon Press, Oxford, UK, 1998).

63. \ J.Hoste, \textit{Transactions AMS} \textbf{297},547 (1986).

64.\ \ \ W.B.R.Lickorish, \textit{An Introduction to Knot Theory}

\ \ \ \ \ \ \ (Springer-Verlag, Berlin, 1997).

65. \ \ W. Massey, \textit{Algebraic Topology: An Introduction}

\ \ \ \ \ \ \ (Springer-Verlag, Berlin, 1967).

66. \ \ C.Adams, \textit{The Knot Book} (W.H.Freeman \&Co., New York, 1994).

67. \ \ D.Cimasoni, \ \textit{Comm.Math.Phys.}\textbf{\ 316}, 99 (2012).

68. \ \ G.Masbaum and A.Vaintrob, \textit{Adv.Math}. \textbf{180}, 765
(2003).

69. \ \ T.Cohran, \textit{Invent.Math}. \textbf{82}, 527 (1985).

70.\ \ \ J.Hoste, \textit{Proceedings AMS} \textbf{95}, 299 (1985).

71. \ \ H.Murakami, Math.Res.Notes of Kobe Univ.\textbf{11},335 (1983).

72. \ \ L.Kauffman, \textit{Cont.Math.} \textbf{44},101 (1985).

73. \ \ L.Guillou and A.Marin, \textit{L'Enseigment Mathematique} \textbf{38}%
, 233 (1992).

74. \ \ G.Burde and H.Zieschang, \textit{Knots} (Walter de Gruyter, New
York, 2003).

75. \ \ T.Ohtsuki, \textit{Quantum Invariants} (World Scientific Publishing
Co.,

\ \ \ \ \ \ \ Singapore, 2002).

76. \ \ C. Venkataraman and W. Miller, \textit{J. Phys. Chem. A}\textbf{\ 108%
}, 3035 (2004).

77. \ \ P.Frampton, \textit{Dual Resonance Models} (W.A.Bejamin,Inc. ,

\ \ \ \ \ \ \ Reading, MA, 1974).

78. \ G. Veneziano, \textit{Nuovo Cim. A} \textbf{57},190 (1968).

79. \ A.Selem and F.Wilczek, hep-ph/0602128.

80. \ S.Chekanov and B.Levchenko, \textit{Phys.Rev.D} \textbf{75}, 014007
(2007).

81. \ J.Sonnenschein and D.Weissman, \textit{JHEP} \textbf{1408}, 013 (2014).

82. \ J.Sonnenschein and D.Weissman, \textit{JHEP} \textbf{1502}, 147 (2015).

83. \ S.Hellerman and I. Swanson, \textit{PRL }\textbf{114},11601 (2015).

84. \ A.Kholodenko, \textit{Int.J.Geom.Methods Mod.Phys}. \textbf{8,}1355
(2011).

85. \ \ Y.Nambu, \textit{Phys.Rev.D} \textbf{10}, 4268 (1974).

86. \ A.Kholodenko and E.Ballard, \textit{Physica A}\textbf{\ 388} (2009)
3024.

87. \ R.Gelca,\textit{\ Theta Functions and Knots},

\ \ \ \ \ \ \ (World Scientific Publishing Co.,Singapore, 2014).

88. \ D.Finkelstein and J.Rubinstein,\ J.Math.Phys.\textbf{9, }1762 (1968).

89. \ S.Krusch and J. Speight, \textit{Comm. Math. Phys.} \textbf{264}, 391
(2006).

90. \ C. Holzhey, F. Wilczek, \textit{Nucl. Phys. B} \textbf{380},447 (1992).

91. \ C.Livingston, \ \textit{Am.Math.Monthly} \textbf{110}, 361(2003).

92. \ H.Moffatt, \textit{Nature} \textbf{347}, 367 (1990).

93. \ J.Hempel, \textit{Topology of 3-manifolds and related topics}

\ \ \ \ \ \ (Prentice-Hall, Englewood Cliffs, N.J.,1961), pp. 207--212.

94. \ R.Kirby, \textit{Invent. Math}. \textbf{45}, 35 (1978).

95. \ D.Rolfsen, \textit{LNM} \textbf{438}, 415 (1975).

96. \ R.Finkelstein,\textit{\ Int.J.Mod.Phys. A} \textbf{29} ,1450092 (2014).

97. \ R.Gompf, A.Stipshicz, \textit{4-Manifolds and Kirby Calculus}

\ \ \ \ \ \ \ (AMS Publishers, Providence, RI, 1999).

98. \ K.Habiro, \textit{Geometry \&Topology} \textbf{10},1285 (2006).

99. \ T.Cohran, S.Friedl and P.Teicher,

\ \ \ \ \ \ \ \textit{Comm. Math.Helv}.\textbf{84}, 617 (2009).

100.\ J.Collins, \textit{Seifert matrices and slice knots},

\ \ \ \ \ \ \ http://www.maths.ed.ac.uk/\symbol{126}s0681349/report.pdf

101.\ J.Levine, \textit{Topology} \textbf{26},45 (1987).

102.\ J.Milnor, \textit{Annals Math}. \textbf{59},177 (1954).

103. \ D.Goldsmith, \textit{Comm.Math.Helvetici }\textbf{54}, 347 (1979).

104. \ C. Griffen, \textit{Math.Scand}.\textbf{45},248 (1979).

105. \ S.Matveev, \textit{Mat. Zametki} \textbf{42}, 268 (1987).

106. \ A.Fomenko and S. Matveev, \textit{Algorithmic and Computer Methods for%
}

\ \ \ \ \ \ \ \ \textit{Three-Manifolds} (Kluwer Academic Publishers,
Boston, 1997).

107. \ C.Lescop, \ \textit{Topology} \textbf{37}, 25 (1998).

108. \ R.Gompf, \textit{Topology} \textbf{25}, 353 (1986).

109. \ M.Freedman and F.Quinn, \textit{The Topology of 4-manifolds}

\ \ \ \ \ \ \ \ (Princeton U.Press, Princeton, NJ 1990).

110. \ V.Krushkal, \textit{\ Annals Math}. \textbf{168}, 675 (2008).

111. \ M.Goussarov, \textit{St. Petersburg Math. J}. \textbf{12}, 569 (2001).

112. \ K.Habiro, \textit{Geometry \& Topology} \textbf{4}, 1 (2000).

113. \ S.Garoufalidis, M.Goussarov and M.Polyak,

\ \ \ \ \ \ \ \ \textit{Geometry \& Topology} \textbf{5},75 (2001).

114. \ \ A.Donald, B.Owens, \textit{Algebraic \& Geometric Topology} \textbf{%
12}, 2069 (2012).

115. \ \ S. Donaldson, \textit{J.Diff. Geom}.\textbf{\ 18}, 279 (1983).

116. \ \ T.Tanaka, \ Topology \& Applications (2015),in press,

\ \ \ \ \ \ \ \ \ http://dx.doi.org/10.1016/j.topol.2015.05.059

117. \ \ J.Meier, \ arXiv: 1401.1161 \ \ 

118. \ \ J.Meier, \textit{Exceptional Seifert fibered surgeries on
Montesinos knots and}

\ \ \ \ \ \ \ \ \ \textit{distinguishing smoothly and topologically doubly
slice knots},

\ \ \ \ \ \ \ \ \ PhD thesis, (Department of Mathematics, U.of Texas at
Austin ,2014)

\ \ \ \ \ \ \ \ \ http://repositories.lib.utexas.edu/handle/2152/24933

119. \ \ M. Atiyah and N.Manton, \textit{Phys.Lett.B} \textbf{222},438
(1989).

120. \ \ N.Manton and P.Sutcliffe, \textit{Topological Solitons }

\ \ \ \ \ \ \ \ (Cambridge U.Press, Cambridge, UK, 2004).

121. \ \ A.Abbas, \textit{Phys.Lett.B} \textbf{593}, 81 (2001).

122. \ \ P.Jennings, Knots and Planar Skyrmions,

\ \ \ \ \ \ \ \ PhD Thesis (Department of physics,Durham University, UK,2015)

\ \ \ \ \ \ \ \ http: //etheses.dur.ac.uk/11161/

123. \ E.Radu and M.Volkov, Phys.Reports \textbf{468}, 101 (2008).

124. \ C.Rohwer and K.M\"{u}ller-Nedebok, \textit{J.Stat.Phys.} \textbf{159}%
, 120 (2015).

125. \ A.Kholodenko and Th.Vilgis, \textit{Phys.Reports} \textbf{298}, 251
(1998).

126. \ A. Kholodenko, \textit{Int.J.Geom.Meth.Mod.Phys}.\textbf{6}, 769
(2009).

127. \ A.Kholodenko, \textit{J. Geom. Phys}., \textbf{59}, 600 (2009).

128. \ Y.Lim, \textit{Math.Res.Lett.} \textbf{6}, 631 (1999).

129. \ P.Kronheimer and T.Mrowka, \textit{Algebraic \&Geom.Topology} \textbf{%
10}, 1715 (2010).

130. \ S.Miller, \textit{Experimental Math}.\textbf{10}, 419 (2005).

131. \ D.Rolfsen, \textit{Pacific.J.Math. }\textbf{110}, 377 (1984).

132. \ D.Rolfsen,\textit{\ Proceedings AMS} \textbf{90},463 (1984).

\bigskip

\bigskip

\bigskip

\bigskip

\bigskip

\bigskip

\bigskip

\bigskip

\bigskip

\bigskip

\bigskip

\medskip

\end{document}